\documentclass[10pt,journal,compsoc]{IEEEtran}
\usepackage{amsmath,amssymb,amsfonts,amsthm}
\usepackage{mathrsfs}
\usepackage{array}
\usepackage{booktabs}
\usepackage{orcidlink}
\usepackage[caption=false,font=normalsize,labelfont=sf,textfont=sf]{subfig}
\usepackage{textcomp}
\usepackage{stfloats}
\usepackage{url}
\usepackage{verbatim}
\usepackage{multirow}
\usepackage{graphicx}
\usepackage{rotate}
\usepackage{amssymb}
\usepackage{cite}
\usepackage{xcolor}
\definecolor{color}{rgb}{0, 0, 1}
\definecolor{color1}{rgb}{1, 0, 0}
\usepackage{bm}
\usepackage[ruled,linesnumbered]{algorithm2e}

\newtheorem{theorem}{Theorem}
\newtheorem{corollary}{Corollary}

\usepackage{algpseudocode}
\makeatletter
\newcommand{\removelatexerror}
{\let\@latex@error\@gobble}
\makeatother
\usepackage{mathtools}
\usepackage{epstopdf}
\usepackage{stfloats}

\begin{document}
	
	\title{Spatiotemporal Non-Uniformity-Aware Online Task Scheduling in Collaborative Edge Computing for Industrial Internet of Things}
	
	\author{
		Yang~Li$^{\orcidlink{0009-0007-1601-633X}}$, Xing~Zhang$^{\orcidlink{0000-0003-4345-6166}}$,~\IEEEmembership{Senior Member,~IEEE,}
        Yukun~Sun, Wenbo~Wang$^{\orcidlink{0000-0002-0911-3189}}$,~\IEEEmembership{Senior Member,~IEEE,} and Bo~Lei$^{\orcidlink{0000-0002-6301-048X}}$
		\thanks{Manuscript received 27 November 2024; revised 17 March 2025 and 21 April 2025; accepted 3 May 2025. This work is supported by the National Science Foundation of China under Grant 62271062, and by the BUPT Excellent Ph.D. Students Foundation under Grant CX20241066. (Corresponding author: Xing Zhang.)}
		\IEEEcompsocitemizethanks{
			\IEEEcompsocthanksitem Yang Li, Xing Zhang, Yukun Sun and Wenbo Wang are with the School of Information and Communications Engineering, Beijing University of Posts and Telecommunications, Beijing 100876, China. E-mail: ly209991@bupt.edu.cn, zhangx@ieee.org, sunyukun@bupt.edu.cn, wbwang@bupt.edu.cn.
                \IEEEcompsocthanksitem  Bo Lei is with Beijing Branch of China Telecom Co., Ltd., Beijing 100032, China. E-mail: leibo@chinatelecom.cn.
                }
	
	}

\IEEEtitleabstractindextext{%
	\begin{abstract}
	\label{abstract}
	Mobile edge computing mitigates the shortcomings of cloud computing caused by unpredictable wide-area network latency and serves as a critical enabling technology for the Industrial Internet of Things (IIoT). Unlike cloud computing, mobile edge networks offer limited and distributed computing resources. As a result, collaborative edge computing emerges as a promising technology that enhances edge networks' service capabilities by integrating computational resources across edge nodes. This paper investigates the task scheduling problem in collaborative edge computing for IIoT, aiming to optimize task processing performance under long-term cost constraints. We propose an online task scheduling algorithm to cope with the spatiotemporal non-uniformity of user request distribution in distributed edge networks. For the spatial non-uniformity of user requests across different factories, we introduce a graph model to guide optimal task scheduling decisions. For the time-varying nature of user request distribution and long-term cost constraints, we apply Lyapunov optimization to decompose the long-term optimization problem into a series of real-time subproblems that do not require prior knowledge of future system states. Given the NP-hard nature of the subproblems, we design a heuristic-based hierarchical optimization approach incorporating enhanced discrete particle swarm and harmonic search algorithms. Finally, an imitation learning-based approach is devised to further accelerate the algorithm’s operation, building upon the initial two algorithms. Comprehensive theoretical analysis and experimental evaluation demonstrate the effectiveness of the proposed schemes.
	\end{abstract}
	
}


\maketitle

\section{Introduction}
\label{Section:Introduction}
\subsection{Research Background and Motivation}

\par \IEEEPARstart{M}{obile} edge computing (MEC) has emerged as a promising architecture for delivering real-time or low-latency services to users in close proximity. This aligns with the Internet of Things (IoT) trend, where emerging applications often require complex computations and low-latency performance, while IoT devices have limited computational capabilities. Moreover, offloading computational tasks to remote cloud processing introduces significant latency due to long backhaul links. Consequently, edge computing has been proposed as a crucial component of IoT, particularly in Industrial Internet of Things (IIoT) applications. For example, it can be used to verify whether workers are wearing protective gear correctly and to analyze equipment status monitoring \cite{1}. Additionally, edge computing addresses privacy concerns in IIoT, such as managing confidential product data that may contain sensitive information.

\par A typical form of mobile edge computing is to equip network access points with storage and computing capabilities, forming edge nodes. This allows computational tasks to be processed at the network's edge, reducing service response time. Compared to cloud computing, edge nodes possess limited computational resources. Meanwhile, the increasing service demand from end devices may overload MEC installations, while cost constraints restrict the expansion of their computing and storage capabilities. These challenges highlight the need for an end-edge-cloud collaboration framework, where end devices primarily offload computation tasks to the nearby MEC server \cite{7,8,9,10}. When computational demand exceeds the capacity of the edge node, some tasks can be further offloaded to the cloud. This collaborative architecture facilitates dynamic task allocation and optimizes computing resource utilization, reducing overall system costs while maintaining quality of service (QoS).

\par While initial research in edge computing has focused on practical topics such as resource allocation, co-scheduling, and latency reduction \cite{3,4,5,6,11,12,13}, certain IIoT applications present unique and fundamentally different challenges. These issues remain largely unexplored but are of significant importance. On one hand, some IIoT applications involve private data that must be processed exclusively by the edge server within the factory \cite{+1in1stMarjor}. An enterprise typically operates multiple factories, each of which deploys an edge computing server. In this context, resource allocation and co-scheduling in edge computing are restricted to horizontal collaboration among edge servers, whereas the traditional end-edge-cloud framework emphasizes vertical collaboration. On the other hand, service requests for IIoT devices across different factories vary over time and are spatially non-uniform \cite{2}. Moreover, enterprises typically operate under long-term cost constraints, and task scheduling among edge nodes occurs over the network, incurring additional overhead. Consequently, in IIoT scenarios that involve processing privacy-sensitive data, task scheduling is confined to horizontal collaboration among edge nodes. It must consider both spatial and temporal collaboration to balance task processing efficiency and long-term operational costs.

\par Nevertheless, satisfying all the aforementioned requirements poses significant challenges for the following reasons.
\begin{enumerate}
\item{Given the time-varying and stochastic nature of the IIoT system state and service requests, horizontal collaboration strategies must be dynamically adjusted among edge nodes. This necessitates online optimization, where long-term cost budgeting relies on statistical data about future network states and service request distributions, which is difficult, if not impossible, to obtain.}
\item{The spatial non-uniformity of service requests and resource allocation across different factories necessitates the careful design of horizontal collaboration among edge nodes to enhance task processing efficiency. However, as the number of factories increases, collaborative decisions must account for the resulting computational complexity, since scalability and real-time feasibility are crucial in IIoT scenarios.}
\item{Spatial and temporal collaboration among edge nodes are interdependent, increasing the complexity of the optimal decision problem and rendering traditional independent optimization methods ineffective.}
\end{enumerate}

\subsection{Related Work}

\par In edge computing networks, edge nodes offer localized, distributed, and limited computational resources. Due to the resource limitations of individual edge nodes and the dynamic, non-uniform distribution of task processing requests, collaborative edge computing has garnered increasing attention from researchers. In this section, we review publications addressing the task scheduling problem in collaborative edge computing networks.

\par Several studies have investigated task allocation between multiple users and multiple edge nodes \cite{15,16,+14in1stMarjor}. In our previous work \cite{15}, we proposed a task offloading framework for multi-user, multi-edge servers aimed at minimizing the average task processing latency by jointly optimizing task scheduling, bandwidth, and computational resource allocation at the edge nodes. Tran \textit{et al.} \cite{16} explored the computational offloading problem involving multiple users and edge nodes. They optimized the weighted sum of task completion time and energy consumption, constrained by the condition that each task is processed on only one edge node. Ma \textit{et al.} \cite{+14in1stMarjor} introduced a truthful combinatorial double auction mechanism for MEC in IIoT. The mechanism was designed to ensure truthfulness and budget balance under locality constraints by optimizing resource allocation and pricing. However, these studies assume that end devices can directly access all edge nodes, making them unsuitable for IIoT scenarios with restricted access ranges. In practice, IIoT devices in a factory can only transmit data to the edge node within the same factory, not to edge nodes in other locations. Consequently, tasks must be redistributed among collaborating edge nodes via the network.

\par Several studies have explored how edge nodes can further offload tasks to collaborative nodes, ensuring acceptable latency and cost \cite{21,22,11}. Lee \textit{et al.} \cite{11} introduced a task scheduling algorithm that prioritizes tasks based on their deadlines and optimizes network flow allocation to minimize the deadline miss ratio of applications. Ebrahimzadeh \textit{et al.} \cite{21} examined collaborative edge computation in FiWi-enhanced HetNets, where Optical Network Units (ONU) are equipped with edge computing servers. The central Optical Line Terminal (OLT), located at a distance from the ONU in the fiber backhaul, is also equipped with computation servers that assist the ONU in further offloading tasks. Hou \textit{et al.} \cite{22} studied task scheduling in collaborative edge computing, using incentives to encourage the allocation of tasks to the optimal computing node for processing. However, these studies focus solely on short-term performance gains, neglecting the effects of fluctuations caused by the time-varying distribution of user requests on long-term utility.

\par Online optimization is a powerful technique for managing dynamic MEC systems with inherent uncertainties while ensuring long-term performance. Several studies have applied deep reinforcement learning (DRL) to MEC resource management and task scheduling optimization \cite{18,19,20,30}. However, the DRL training process may encounter policy instability, over-exploration, or oscillations, leading to unstable system performance or even failure to meet QoS requirements \cite{+2in1stMarjor}. Additionally, as a black-box model, DRL lacks strict performance guarantees. To address these challenges, Lyapunov-based online optimization offers a viable alternative. The Lyapunov-based online optimization approach offers an efficient, stable, and interpretable framework for MEC resource management, optimizing performance while ensuring long-term system stability \cite{+3in1stMarjor}. Existing research primarily focuses on single-timescale online decision-making based on Lyapunov optimization \cite{+3in1stMarjor, +4in1stMarjor, +5in1stMarjor } or extends to two-timescale joint online decision-making \cite{+6in1stMarjor, +7in1stMarjor, +8in1stMarjor }. However, these studies focus solely on optimization along the time dimension and cannot be directly applied to the problem addressed in this paper. Due to the spatiotemporal non-uniformity of service requests from IIoT devices in different factories, the optimization in the time and spatial dimensions are inherently coupled.

\par In recent years, the use of graph models in network optimization has garnered significant attention, particularly in areas such as routing optimization \cite{+9in1stMarjor, +10in1stMarjor}  and resource allocation \cite{20, +12in1stMarjor, +13in1stMarjor}. Studies in \cite{+9in1stMarjor} and \cite{+10in1stMarjor} represented the network as a graph and employed graph theory algorithms to identify the shortest paths. In \cite{20}, services were modeled as directed acyclic graphs (DAGs) to optimize task partitioning. Studies in \cite{+12in1stMarjor} and \cite{+13in1stMarjor} represented the relationships between tasks and computing nodes as bipartite graphs and applied graph theory algorithms to solve the optimal matching problem. However, these studies also do not consider the spatiotemporal non-uniformity of service request distribution and suffer from high computational costs in large-scale network environments. In contrast to the above research, we leverage graph models to guide task scheduling, narrow the search space for decision-making, and employ a graph neural network for imitation learning to further reduce algorithm runtime.

\begin{table*}[t]
\centering
\caption{A Table Comparing Our Work With The Existing Studies}
\label{tab0}
\Large
\resizebox{\textwidth}{!}{%
\begin{tabular}{cccccc}
\midrule[1.5pt]
\textbf{Reference}                 & \textbf{Method}                                                                                                         & \textbf{Task Scheduling Model}                                                                                                                                  & \textbf{\begin{tabular}[c]{@{}c@{}}Consideration of \\ Time-Varying Nature \end{tabular}} & \textbf{\begin{tabular}[c]{@{}c@{}}Consideration of  \\ Spatial Non-Uniformity \end{tabular}} & \textbf{\begin{tabular}[c]{@{}c@{}}Prototype Implementation \\ and Validation\end{tabular}} \\ \midrule[1.5pt]
\cite{15,16}                       & \begin{tabular}[c]{@{}c@{}}Low-complexity heuristic \\ approach\end{tabular}                                            & \begin{tabular}[c]{@{}c@{}}Users directly offload \\ tasks to the corresponding\\  MEC servers.\end{tabular}                                                    & No                                                                                                                                & No                                                                                                                                  & No                                                                                          \\ \hline
\cite{+14in1stMarjor}              & \begin{tabular}[c]{@{}c@{}}Combinatorial double \\ auction-based algorithm\end{tabular}                                 & \begin{tabular}[c]{@{}c@{}}Users directly offload \\ tasks to the corresponding \\ MEC servers.\end{tabular}                                                    & No                                                                                                                                & No                                                                                                                                  & No                                                                                          \\ \hline
\cite{21}                          & \begin{tabular}[c]{@{}c@{}}Self-organization based \\ mechanism\end{tabular}                                            & \begin{tabular}[c]{@{}c@{}}Tasks are redistributed \\ among edge nodes via the \\ network.\end{tabular}                                                         & No                                                                                                                                & Yes                                                                                                                                 & No                                                                                          \\ \hline
\cite{22}                          & Incentive-based approach                                                                                                & \begin{tabular}[c]{@{}c@{}}Tasks are redistributed \\ among edge nodes and \\ auxiliary devices.\end{tabular}                                                   & No                                                                                                                                & Yes                                                                                                                                 & No                                                                                          \\ \hline
\cite{11}                          & \begin{tabular}[c]{@{}c@{}}Low-complexity heuristic \\ approach\end{tabular}                                            & \begin{tabular}[c]{@{}c@{}}Tasks are redistributed \\ among edge nodes via the \\ network.\end{tabular}                                                         & No                                                                                                                                & Yes                                                                                                                                 & No                                                                                          \\ \hline
\cite{18,19,20,30}                 & DRL-based algorithm                                                                                                     & \begin{tabular}[c]{@{}c@{}}Tasks are divided into \\ multiple subtasks and \\ scheduled in parallel \\ to multiple edge \\ servers for processing.\end{tabular} & Yes                                                                                                                               & No                                                                                                                                  & No                                                                                          \\ \hline
\cite{+3in1stMarjor}               & \begin{tabular}[c]{@{}c@{}}Lyapunov-based single-\\ timescale online \\ optimization algorithm\end{tabular}             & \begin{tabular}[c]{@{}c@{}}Tasks are redistributed \\ among edge nodes via \\ the network.\end{tabular}                                                         & Yes                                                                                                                               & No                                                                                                                                  & No                                                                                          \\ \hline
\cite{+4in1stMarjor,+5in1stMarjor} & \begin{tabular}[c]{@{}c@{}}Lyapunov-based single-\\ timescale online \\ optimization algorithm\end{tabular}             & \begin{tabular}[c]{@{}c@{}}Tasks are redistributed \\ between edge nodes and \\ the cloud.\end{tabular}                                                         & Yes                                                                                                                               & No                                                                                                                                  & No                                                                                          \\ \hline
\cite{+6in1stMarjor,+7in1stMarjor} & \begin{tabular}[c]{@{}c@{}}Lyapunov-based two-\\ timescale online \\ optimization algorithm\end{tabular}                & \begin{tabular}[c]{@{}c@{}}Users directly offload \\ tasks to the corresponding \\ MEC servers.\end{tabular}                                                    & Yes                                                                                                                               & No                                                                                                                                  & No                                                                                          \\ \hline
\cite{+8in1stMarjor}               & \begin{tabular}[c]{@{}c@{}}Lyapunov-based two-\\ timescale online \\ optimization algorithm\end{tabular}                & \begin{tabular}[c]{@{}c@{}}Tasks are redistributed \\ among edge nodes via \\ the network.\end{tabular}                                                         & Yes                                                                                                                               & No                                                                                                                                  & No                                                                                          \\ \hline
\cite{+9in1stMarjor, +10in1stMarjor}               & \begin{tabular}[c]{@{}c@{}}Shortest path algorithm \end{tabular}                & \begin{tabular}[c]{@{}c@{}}Tasks are transmitted  \\ according to the specified \\ path. \end{tabular}                                                         & No                                                                                                                               & No                                                                                                                                  & No                                                                                          \\ \hline
\cite{+12in1stMarjor, +13in1stMarjor}               & \begin{tabular}[c]{@{}c@{}}Bipartite graph matching\\ algorithm\end{tabular}                & \begin{tabular}[c]{@{}c@{}}Users directly transmit \\  tasks to the computing  \\ nodes.\end{tabular}                                                         & No                                                                                                                               & No                                                                                                                                  & No                                                                                          \\ \hline
Proposed work                      & \begin{tabular}[c]{@{}c@{}}A approach integrates \\ Lyapunov optimization \\ with the graph-based \\ model\end{tabular} & \begin{tabular}[c]{@{}c@{}}Tasks are redistributed \\ among edge nodes via \\ the network.\end{tabular}                                                         & Yes                                                                                                                               & Yes                                                                                                                                 & Yes                                                                                         \\ \midrule[1.5pt]
\end{tabular}%
}
\end{table*}

\par In summary, different from all existing work (as demonstrated in Table \ref{tab0}), this paper proposes a novel spatiotemporal non-uniformity-aware online task scheduling scheme for collaborative edge computing in IIoT. The proposed scheme integrates Lyapunov optimization with a graph-based model to effectively handle the time-varying and spatially non-uniform distribution of service requests and resources. Furthermore, none of the studies discussed in this subsection have demonstrated their superior performance in real engineering applications, as they are challenging to implement in practice. In this paper, we propose a generic edge node collaboration scheme and develop a prototype testbed using Kubernetes, providing a real-world application case that verifies the superior performance of the proposed scheme.

\subsection{Contribution and Organization}
\par The main contributions of this paper are summarized as follows:
\begin{itemize}
	\item Dynamic task scheduling in the collaborative edge computing network is formulated as a stochastic optimization problem, aiming to optimize long-term average task processing delay under the constraint of long-term operational cost. This problem is NP-hard.
 
	\item The Lyapunov optimization is employed to incorporate the long-term task scheduling cost constraint into real-time optimization, allowing task scheduling decisions to be made online without requiring any prior knowledge of future information (e.g., changes in the distribution of user requests).
 
	\item In terms of the problem derived from the Lyapunov optimization, a graph model is introduced to guide optimal task scheduling decisions, and a two-stage heuristic algorithm is developed.
 
	\item  To further reduce algorithm execution time, an imitation learning-based scheme is developed based on the previous algorithms. Finally, the performance of the proposed algorithm is demonstrated through theoretical analysis, numerical simulations, and engineering applications.
\end{itemize}

The remainder of the paper is structured as follows. Section \ref{Section:System Model} introduces the system model. Section \ref{Section:Problem Formulation and Analysis} formulates and analyzes the task scheduling problem in collaborative edge computing networks, with the goal of minimizing the long-term average task processing delay, subject to the constraint of long-term operational cost. Section \ref{Section:Online Task Scheduling Algorithm} proposes an online task-scheduling framework along with the algorithms implemented within it. Section \ref{Section:Theoretical Analysis} analyzes the complexity and convergence of the proposed algorithms. Section \ref{Section:Performance Evaluation} validates the feasibility and superiority of the proposed algorithms through numerical simulations and practical engineering applications\footnote{The code is available at https://github.com/CPNGroup/Spatial\-Temporal-Non-Uniformity-Aware-Online-Task-Scheduling.}. Section \ref{Section:Conclusion} concludes the paper.

\section{System Model}
\label{Section:System Model}

\par In this section, we first present a comprehensive system description. Subsequently, we detail the considered models, including service request scheduling, response latency, and task processing overhead.

\subsection{System Description}
\label{SubSection:System Description}

\begin{figure}[t]
	\centering	\includegraphics[width=\linewidth,scale=1.00]{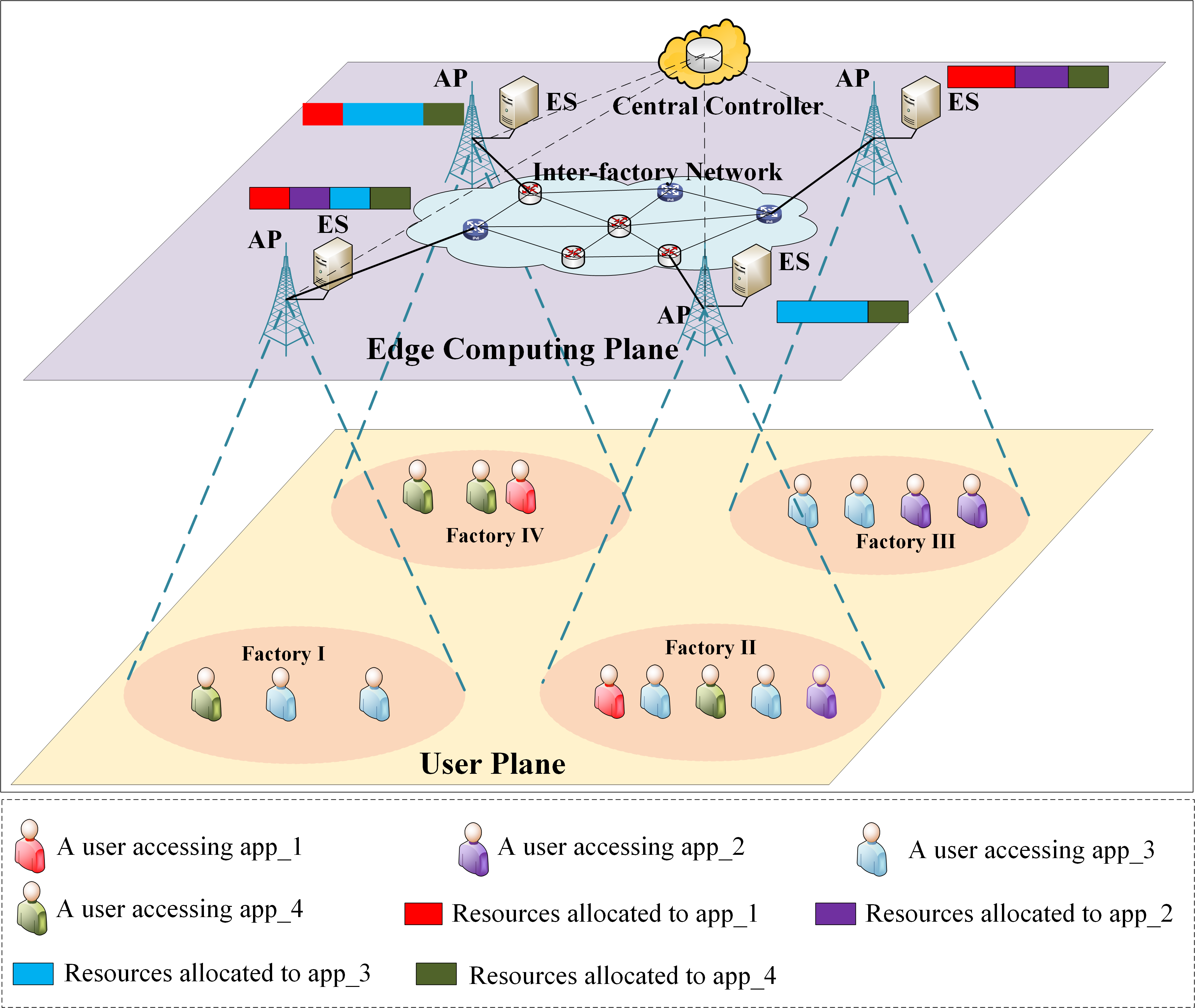}
	\caption{Illustration of the network architecture for collaborative edge computing in the IIoT}
	\label{Figure:System Model}
\end{figure}

\par As illustrated in Fig. \ref{Figure:System Model}, we consider an enterprise with multiple factories located in different geographical regions. For privacy and latency requirements, each factory is equipped with an edge server (ES) at its access point (AP). The APs in all factories are interconnected through the inter-factory network to enable mutual access and task scheduling. Due to the presence of multiple tasks in the factories, services need to be deployed on each ES (as represented by app\_1, 2, 3, 4 in Fig. \ref{Figure:System Model}), and corresponding resources must be allocated. Resource allocation decisions can be made offline, based on the expected task processing requirements of each factory \cite{26}, or online, dynamically adapting to the request distribution of each ES \cite{27,28,29}. Notably, due to the differing time sensitivities of task scheduling and resource allocation, executing these decisions simultaneously would delay task scheduling, violating real-time task requirements. Moreover, this approach would cause the system to orchestrate resources too frequently, increasing operational costs and system instability \cite{30}. Consequently, resource allocation typically operates on a larger timescale than task scheduling. Specifically, resource allocation for each service is determined over a long timescale, and task scheduling occurs over a short timescale \cite{+15in1stMarjor}. This paper focuses on dynamic task scheduling strategies, assuming that service deployment and resource allocation are completed before task scheduling decisions, aligning with real-world engineering practices.

\par The IIoT system, as depicted in Fig. \ref{Figure:System Model}, can be divided into three parts: the user plane, the edge computing plane, and the central controller. The user plane consists of IIoT devices distributed across different factories, which can initiate requests for different services to the APs of their respective factories. The edge computing plane includes APs, ESs, and the inter-factory network. APs execute task scheduling strategies for different services based on control signals from the central controller. The inter-factory network enables task transmission among factories. ESs hand over received tasks to the services deployed on them for processing. The central controller senses the request distribution and system status of each factory, dynamically adjusts task scheduling strategies for different services, and sends them to the APs for execution.

\par
\begin{figure*}[htbp]
	\centering	\includegraphics[width=\linewidth,scale=1.00]{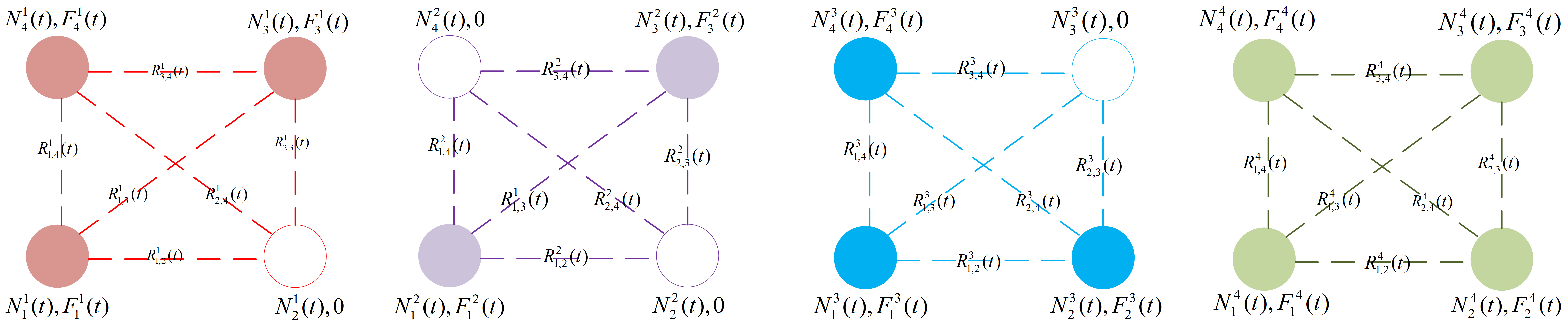}
	\caption{Computing network graph for each service corresponding to Fig. \ref{Figure:System Model} during the $t$th time slot. (a) Computing network graph of app\_1. (b) Computing network graph of app\_2. (c) Computing network graph of app\_3. (d) Computing network graph of app\_4. Solid and hollow circles represent cases where $F_{m}^{k}(t)>0$ and $F_{m}^{k}(t)=0$, respectively.}
	\label{Figure:CNCgraph}
\end{figure*}

 For simplicity in the following discussion, we unify the notation for ESs, APs, and factories, collectively denoting them as $\mathcal{M}=\{1,2,\ldots, M\}$. All deployed services are represented as $\mathcal{K}=\{1,2,\ldots,K\}$. Given the dynamic nature of service request distribution, the system operates in time slots, with time discretized into frames $\mathcal{T}=\{1,2,\ldots T\}$. Notably, the division of time frames may vary across services, depending on their maximum tolerable delay. To describe the spatiotemporal characteristics of request distribution in the user plane and resource allocation in the edge computing plane, denote $N_{m}^{k}(t)$ as the number of $k$th service requests arriving at the $m$th AP at the beginning of the $t$th time slot. Let $F_m^k(t)$ represent the computational resources allocated by ES $m$ to the $k$th service at the start of the $t$th time slot. Let $R_{m,m^{\prime}}^k(t)$ denote the bandwidth allocated for transferring requests of the $k$th service from AP $m$ to ES $m^{\prime}$ during the $t$th time slot, where the first and second subscripts represent the AP and ES indices, respectively. In practice, bandwidth allocation in wired networks based on service demand is common. Since each AP is directly connected to its co-located ES via fiber, we assume $R_{m,m}^k(t) = \infty$, i.e., the transmission delay between AP $m$ and ES $m$ is negligible. Using the request distribution from the user side and the resource allocation from the provisioning side, we can model the state of each service at the $t$th time slot as a computing network graph, as shown in Fig. \ref{Figure:CNCgraph}.

Fig. \ref{Figure:CNCgraph} shows that each service's computational network graph is independent, allowing task scheduling decisions to be made separately for each service \cite{+1}. To clarify, we will analyze the $k$th service as an example in the following discussion. The important notations used in the rest of this paper are summarized in Table \ref{Tab:Major Notions}.

\begin{table}[h]
    \centering
    \renewcommand{\arraystretch}{1.3}
    \caption{Major Notions}
    \label{Tab:Major Notions}
    \resizebox{\columnwidth}{!}{%
    \begin{tabular}[c]{p{0.1cm}|cl}
        \toprule[1.5pt]
        & {\textbf{Symbols}} & {\textbf{Definition}} \\
        \toprule[1pt]
        \multirow{18}{*}{\rotatebox{90}{\textbf{System Model}}}
        & $\mathcal{M}$ & Set of ESs/APs/factories \\
        & $\mathcal{K}$ & Set of services \\
        & $\mathcal{T}$ & Set of time slots \\
        & $N_m^k(t)$ & Number of requests for app\_k arriving at the $m$th AP at the start \\& &of the $t$th time slot \\
        & $N_{m,m^{\prime}}^k(t)$ & Number of requests for app\_k arriving at the $m$th AP at the start \\& &of the $t$th time slot scheduled for processing on ES $m^{\prime}$. \\
        & $F_m^k(t)$ &  Computing resources allocated by the $m$th ES to app\_k during\\& & the $t$th time slot. \\
        & $R_{m,m^{\prime}}^k(t)$ & Bandwidth allocated for scheduling requests of app\_k from AP\\& & $m$ to ES $m^{\prime}$ during the $t$th time slot. \\
        & $T_k(t)$ & Average response delay of all requests for app\_k in the $t$th time \\& &slot. \\
        & $E_k(t)$ & Task processing cost of all requests for app\_k in the $t$th time slot \\
        & $E_k^{avg}$ & Long-term average task processing cost budget for app\_k \\
        & $\boldsymbol{\mathcal{N}}_k$ & Online task scheduling decisions for app\_k \\
        & $\mathcal{N}_k(t)$ & Task scheduling decision for app\_k in the $t$th time slot. \\
        \midrule[1.5pt]
        \multirow{6}{*}{\rotatebox{90}{\textbf{Algorithm}}}
        & $Q_k(t)$ & Virtual control queue length for app\_k at the $t$th time slot. \\
        & $V$ & Weight parameter for balancing the trade-off between the task \\& &response latency performance and the task processing cost budget \\& &violations\\
        & $\mathcal{O}$ & The set of factory/region index for isolated nodes \\
        & $\mathcal{P}$ & The set of factory/region index for source nodes \\
        & $\mathcal{Q}$ & The set of factory/region index for sink nodes \\
        & $NI$ & Maximum number of iterations for Algorithm 2 \\
        & $I_{max}$ & Maximum number of iterations for Algorithm 3 \\
        \bottomrule[1.5pt]
    \end{tabular}
    }
\end{table}

\subsection{Service Request Scheduling Model}
\label{SubSection:Service Request Scheduling Model}

\begin{figure}[t]
	\centering	\includegraphics[width=\linewidth,scale=1.00]{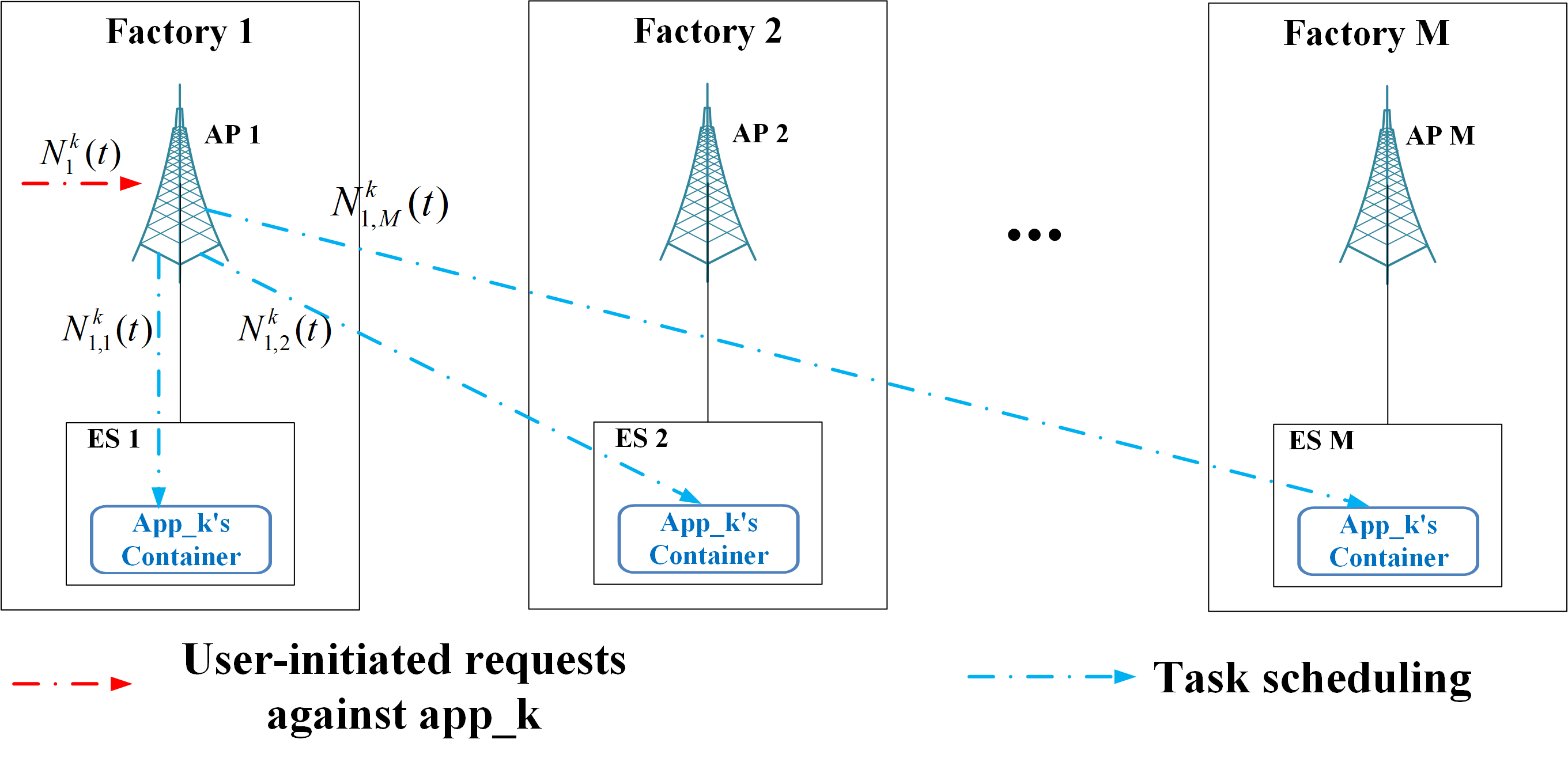}
	\caption{Illustration of the processing flow for requests against the $k$th service initiated by IIoT devices in Factory 1}
	\label{Figure:Processing Flow}
\end{figure}

\par Fig. \ref{Figure:Processing Flow} shows the processing flow of requests for the $k$th service initiated by IIoT devices in Factory 1 during the $t$th time slot. These requests first reach the AP in Factory 1, where the task scheduler for the $k$th service allocates them to ESs hosting the $k$th service, following the central controller's dispatch policy. Let $N_m^k (t)$ denote the number of requests for the $k$th service initiated by the IIoT devices in Factory $m$ during the $t$th time slot. The number of these requests scheduled for processing on ES $m^{\prime}\in\mathcal{M}$ is represented by $N_{m,m^{\prime}}^k (t)$, we have:
\begin{equation}
\label{II-a1}
\sum_{m^{\prime}\in\mathcal{M}}N_{m,m^{\prime}}^k(t)=N_m^k(t),
\end{equation}
\begin{equation}
\label{II-a2}
\begin{cases}N_{m,m'}^k(t)=0&if \quad F_{m'}^k(t)=0\\N_{m,m'}^k(t)\geq0&if \quad F_{m'}^k(t)>0\end{cases}.
\end{equation}

\par Equation (\ref{II-a2}) indicates that if ES $m^{\prime}$ does not allocate resources for the $k$th service, it will not process any requests for that service. Building on the previously defined task scheduling decisions, the response delay and processing overhead resulting from these decisions can be quantified.

\subsection{Response Latency Model}
\label{SubSection:Response Latency Model}

\par In the considered scenarios, a task's response latency is affected by both scheduling and computation latencies. We exclude task offloading latency and result return latency from consideration, as task scheduling decisions do not impact offloading latency, and the result size is sufficiently small for its return delay to be negligible \cite{20}. Since our focus is on the average response latency of all requests for the $k$th service during each time slot, we begin by analyzing the total latency of all requests for the $k$th service in the $t$th time slot.
\begin{itemize}
	\item \textbf{\textit{Scheduling Latency}}: Let $L_k$ represent the size of a single request for the $k$th service. Based on the task scheduling decision, we can calculate the total latency of all requests for the $k$th service transmitted from AP $m$ to ES $m^{\prime}$ in the $t$th time slot as
    \begin{equation}
    \label{II-b1}
    T_{m,m^{\prime}}^{k,tran}(t)=\frac{N_{m,m^{\prime}}^k(t)L_k}{R_{m,m^{\prime}}^k(t)},
    \end{equation}
    where the numerator represents the total data size of all requests for the $k$th service scheduled from AP $m$ to ES $m^{\prime}$ in the $t$th time slot. Thus, the total scheduling delay of all requests for the $k$th service in the $t$th time slot can be expressed as
    \begin{equation}
    \label{II-b2}
    T_k^{tran}(t)=\sum_{m\in\mathcal{M}}\sum_{m^{\prime}\in\mathcal{M}\backslash\{m\}}T_{m,m^{\prime}}^{k,tran}(t),
    \end{equation}
    where the outer summation runs over all APs, and the inner summation is performed over all ESs communicating with the current AP through the inter-factory network.

	\item \textbf{\textit{Computation Latency}}: Let $C_k$ represent the number of CPU cycles needed to process a single request for the $k$th service. The total computation time of all tasks for the $k$th service on ES $m$ in the $t$th time slot can be calculated as
    \begin{equation}
    \label{II-b3}
    T_m^{k,comp}(t)=\frac{\sum_{m^{\prime}\in\mathcal{M}}N_{m^{\prime},m}^k(t)C_k}{F_m^k(t)},
    \end{equation}
    where the numerator represents the total CPU cycles required to process all requests for the $k$th service that are dispatched from all APs to ES $m$. Thus, the total computation delay of all tasks for the $k$th service in the $t$th time slot can be expressed as
    \begin{equation}
    \label{II-b3+}
    T_k^{comp}(t)=\sum_{m\in\mathcal{M}}T_m^{k,comp}(t).
    \end{equation}

\end{itemize}

Based on the above analyses, the average response delay of all requests for the $k$th service in the $t$th time slot can be calculated as
\begin{equation}
\label{II-b4}
T_k(t)=\frac{T_k^{tran}(t)+T_k^{comp}(t)}{\Sigma_{m\in\mathcal{M}}N_m^k(t)}.
\end{equation}
Specifically, the numerator represents the total response delay of all requests for the $k $th service in the $t $th time slot, and the denominator denotes the total number of requests for the $k $th service initiated by all IIoT devices in the $t $th time slot.

\subsection{Task Processing Cost Model}
\label{SubSection:Task Scheduling Cost Model}

\par Optimizing the task scheduling policy among ESs can reduce the average response latency of services. However, task scheduling introduces additional operational costs. For instance, transferring data across regions raises the energy consumption of network devices like routers and switches. Additionally, processing tasks on edge nodes also consume energy. Therefore, consistent with the previous subsection, we categorize task processing overhead into scheduling cost and computation cost.
\begin{itemize}
	\item \textbf{\textit{Scheduling Cost}:} To model the operational cost of task scheduling, let $E_{m,m^{\prime}}^k$ represent the cost of scheduling a single request for the $k$th service from AP $m$ to ES $m^{\prime}$, where the first and second subscripts represent the AP and ES indices, respectively. Without loss of generality, assume $E_{m,m}^k=0$, i.e., the transmission cost from AP $m$ to ES $m$ is negligible. Given the task scheduling decision for the $k$th service in the $t$th time slot, the task scheduling cost for the $k$th service in the $t$th time slot can be calculated as
    \begin{equation}
    \label{II-c1}
    E_k^{tran}(t)=\sum_{m\in\mathcal{M}}\sum_{m^{\prime}\in\mathcal{M}\backslash\{m\}}N_{m,m^{\prime}}^k(t)E_{m,m^{\prime}}^k, 
    \end{equation}
    where the outer summation runs over all APs, and the inner summation is performed over all ESs communicating with the current AP through the inter-factory network.
	
    \item \textbf{\textit{Computation Cost}}: Assuming that all requests for a given service processed on an edge node share the computational resources occupied by that service equally, the computational resources allocated to each request for the $k$th service on ES $m $ can be expressed as
    \begin{equation}
    \label{II-c1+}
    f_m^k(t) = \frac{{F_m^k(t)}}{\sum_{m^{\prime}\in\mathcal{M}}N_{m^{\prime},m}^{k}(t)},
    \end{equation}
    where the denominator represents the total number of requests for the $k$th service dispatched from all APs to ES $m$ for processing. Let $k_m $ represent the computational energy efficiency factor of ES $m $. According to \cite{+7in1stMarjor} and \cite{24}, the total computational energy consumption of all tasks for the $k$th service processed on ES $m $ during the $t $th time slot can be calculated as
    \begin{equation}
    \label{II-c2}
    E_m^{k,comp}(t)=\sum_{m^{\prime}\in\mathcal{M}}N_{m^{\prime},m}^{k}(t)k_m{f_m^k(t)}^2C_{k},
    \end{equation}
   where the summation runs over all APs. Therefore, the total computational energy consumption of all tasks for the $k$th service in the $t$th time slot can be calculated as
    \begin{equation}
    \label{II-c3}
    E_k^{comp}(t) =\sum_{m\in \mathcal{M}}E_m^{k,comp}(t).
    \end{equation}

\end{itemize}

Based on the above analyses, the total task processing cost of all requests for the $k$th service in the $t$th time slot can be calculated as
\begin{equation}
\label{II-c4}
E_k(t) = E_k^{tran}(t) +E_k^{comp}(t).
\end{equation}

\par On one hand, user requests fluctuate over time, meaning the number of requests for the $k$th service received by each AP changes at every time slot. On the other hand, the distribution of requests for the $k$th service across different factories is uneven. Therefore, to minimize the average response latency, the task scheduling policy must be dynamically adjusted at each time slot. However, frequent task scheduling leads to high operational costs. Thus, the key question is how to efficiently balance the trade-off between cost and performance.

\section{Problem Formulation and Analysis}
\label{Section:Problem Formulation and Analysis}
\subsection{Problem Formulation}
\label{SubSection:Problem Formulation}

\par Considering the fact that enterprises typically operate under the long-term cost budget, we formulate the problem as optimizing task processing performance over the long term within the long-term cost constraint. Specifically, we define $E_k^{avg}$ as the long-term average task processing cost budget for the $k$th service over $T$ time slots ($T \rightarrow \infty$), which satisfies 
\begin{equation}
\label{III-a1}
\mathop {\emph{lim} }\limits_{T\to\infty}\frac1T\sum_{t=1}^TE_k(t)\leq E_k^{avg}.
\end{equation}

\par Our goal is to minimize the long-term average request response latency, which can be formulated as the following stochastic optimization problem:
\begin{equation}
\begin{aligned}\label{P1}
\mathcal{P}_1:&\mathop {\emph{min} }\limits_{\boldsymbol{\mathcal{N}}_k
} \mathop {\emph{lim} }\limits_{T\to\infty} \;\frac{1}{T}\sum_{t=1}^{T} T_{k}(t) \\
\mbox{s.t.}\;&(a): (\ref{II-a1}),(\ref{II-a2}),(\ref{II-b4}),(\ref{II-c4}),(\ref{III-a1}),\\
&(b): N_{m,m^{\prime}}^k (t) \in \mathbb{N},\quad \forall m,m^{\prime}\in\mathcal{M},
\end{aligned}
\end{equation}
where $\mathbb{N}$ denotes the set of non-negative integers, $\boldsymbol{\mathcal{N}}_k = \{\mathcal{N}_k(t)| t =1, 2, \cdots, T\}$, and $\mathcal{N}_k(t)=\{N_{m,m^{\prime}}^k(t)| m,m^{\prime}\in\mathcal{M}\}$. 

\par The primary challenge in addressing problem $\mathcal{P}_1$ lies in the requirement for future system information (i.e., the evolution of request distribution and resource allocation for the $k$th service over time) to enable the decision controller to make globally optimal task scheduling decisions in each time slot. Unfortunately, prediction-based methods often fail to accurately capture the state of a time-varying system, whereas queuing-theory-based methods depend on a priori information (e.g., task arrival and service rates) and assume specific distributional patterns for request arrivals and departures, which are frequently difficult to satisfy. Fortunately, in queuing theory, the long-term task processing cost budget constraint (\ref{III-a1}) in this optimization problem can be interpreted as a form of queue stability control. Moreover, the Lyapunov optimization technique provides an efficient approach to decouple the long-term problem. It does not require any prior system information and can maintain queue stability in an online manner. Therefore, we employ Lyapunov optimization theory to transform the original problem into a series of real-time minimization problems, leading to the development of an online task scheduling algorithm.

\subsection{Graph Model and Optimal Solution Analysis}
\label{SubSection:Optimal Solution Analysis}
\par To enhance the understanding of the problem, we model the system state and task scheduling policy using time-varying graph models, which are described separately below.

\subsubsection{System State Graph Model}
\label{SubSubSection:System State Graph Model}

\par Based on the system model in Section \ref{Section:System Model} and the problem description in the previous subsection, the system state of the $ k $th service can be represented as a time-varying complete undirected graph, denoted by $\mathcal{G}_t^k=\{\mathcal{V}_t^k,\mathcal{E}_t^k\}$, where:
\begin{itemize}
\item{$ \mathcal{V}_t^k $ denotes the set of vertices representing the edge nodes. Each vertex $ m \in \mathcal{V}_t^k $ is associated with two attributes: $ N_m^k(t) $ and $ F_m^k(t) $, as illustrated in Fig. \ref{Figure:CNCgraph}.}
\item{$ \mathcal{E}_t^k $ denotes the set of edges, defined as $ \mathcal{E}_t^k=\{e_{m,m^{\prime}}^k \mid m,m^{\prime} \in \mathcal{V}_t^k\} $. Each edge $ e_{m,m^{\prime}}^k \in \mathcal{E}_t^k $ is associated with two attributes: $ R_{m,m^{\prime}}^k(t) $ and $ E_{m,m^{\prime}}^k $.}
\item{The graph $ \mathcal{G}_t^k $ also possesses two global properties: $ E_{k}^{avg} $ and $ \boldsymbol{E}_{k}^{t-}=\{E_k(l) \mid l=1,2,\dots,t-1\} $, where $ \boldsymbol{E}_{k}^{t-} $ represents the known past task processing cost of the $ k $th service up to the $t$th time slot.}
\end{itemize}

\subsubsection{Task Scheduling Policy Graph Model}
\label{SubSubSection:Task Scheduling Policy Graph Model}
\begin{figure}[t]
\centerline{\includegraphics[width=0.55\columnwidth]{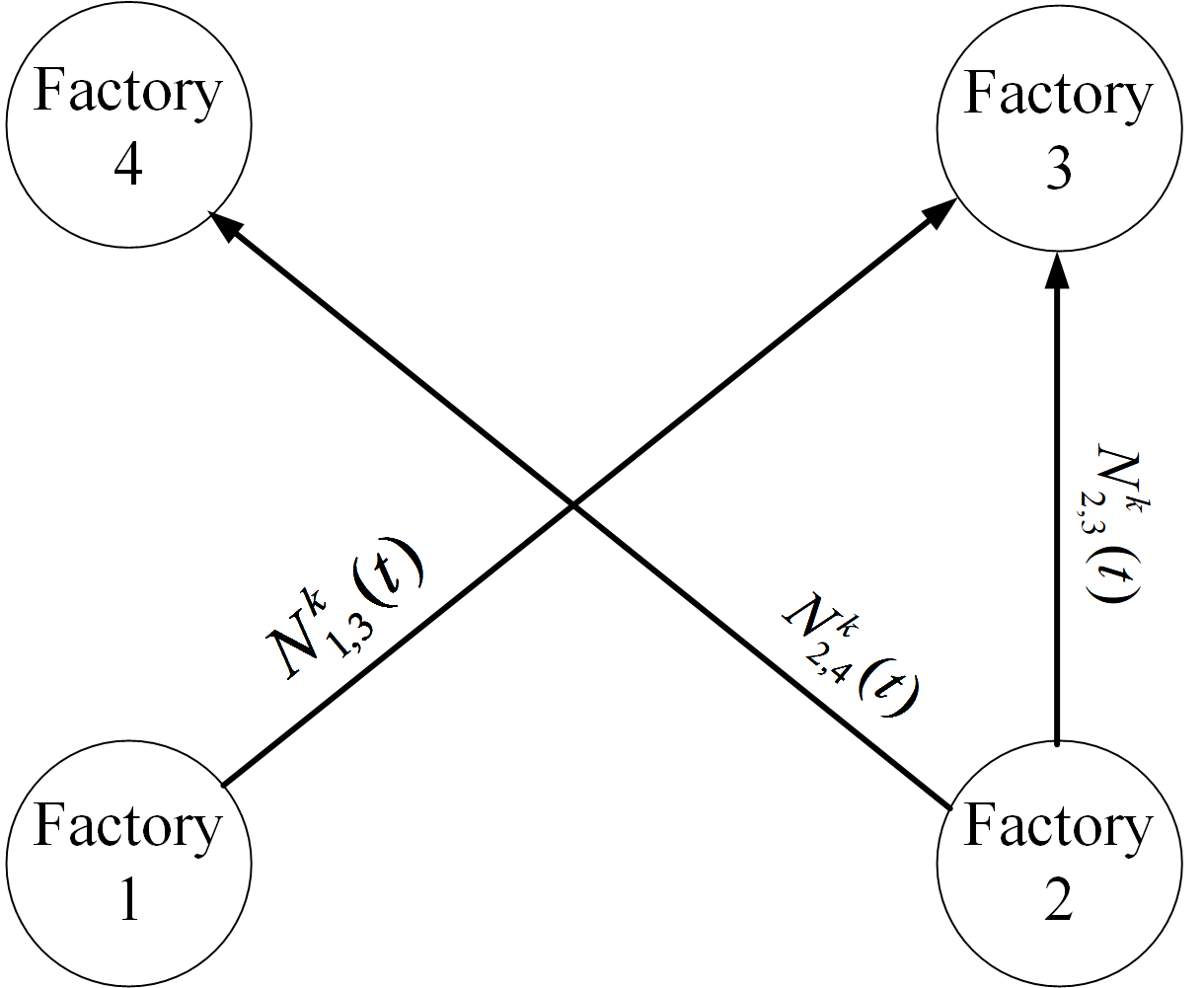}}
    \caption{An example of a weighted directed graph for the task scheduling decision of the $k$th service at the $t$th time slot. This example shows that, in the current time slot, the AP of factory 2 will send $N_{2,3}^k(t)$ requests for $app\_k$ to the ES of factory 3 and $N_{2,4}^k(t)$ requests to the ES of factory 4. The AP of factory 1 will send $N_{1,3}^k(t)$ requests for $app\_k$ to the ES of factory 3.}
\label{Fig:decision example}
\end{figure}

\par Based on the system model in Section \ref{Section:System Model}, the variables to be optimized in $\mathcal{P}_1$ can be modeled as a time-varying weighted directed graph. Fig. \ref{Fig:decision example} illustrates an example of a weighted directed graph for the task scheduling decision of the $k$th service at the $t$th time slot. Each directed edge represents cross-factory task scheduling, and the weight corresponds to the number of forwarded requests. Each node corresponds to a factory or an edge node.

\subsubsection{Optimal Solution Analysis}
\label{SubSubSection:Optimal Solution Analysis}

\par Based on the above description, the following theorem can be stated:

\begin{theorem}
For the $k$th service, the optimal task scheduling decision in any time slot forms a weighted directed acyclic graph that satisfies the following conditions:
\begin{enumerate}
\item{The graph does not contain intermediate nodes, meaning only source nodes, sink nodes, and isolated nodes exist.}
\item{Nodes in set $\{m|F_m^k (t)=0\}$ are source nodes.}
\item{For any ES $m$ corresponding to a source node, if $F_m^k (t)=0$, the sum of the weights of all outgoing edges equals $N_m^k (t)$. If $F_m^k (t)>0$, the sum is less than or equal to $N_m^k (t)$.}
\item{A sink node must exist whenever there is a source node.}
\end{enumerate}
\end{theorem}

\begin{proof}
The detailed proof is given in Appendix A of the supplemental material.
\end{proof}

\section{Online Task Scheduling Algorithm}
\label{Section:Online Task Scheduling Algorithm}

\par In this section, we propose a novel framework for making online task scheduling decisions that are aware of spatiotemporal non-uniformity. To solve $\mathcal{P}_1$, we first transform the original problem into a queue stability control problem using Lyapunov optimization.

\subsection{Problem Transformation}
\label{SubSection:Problem Transformation}

\subsubsection{Construct a Virtual Queue}

\par A key concept in Lyapunov optimization is to decompose the long-term optimization metric with long-term constraints into per-time-slot optimization. This allows for direct optimization of the objective function in each time slot, while adhering to the necessary constraints of that slot, such as maintaining virtual queue stability. Therefore, for the constraint $lim_{T\to\infty}\frac1T\sum_{t=1}^TE_k(t)\leq E_k^{avg}$, we define a virtual queue $Q_k(t)$ with an initial value of 0:
\begin{equation}
\label{IV-a1}
Q_k(t+1)=\begin{cases}0,&t=0\\max\{Q_k(t)+E_k(t)-E_k^{a\nu g},0\},&t=1,2,\ldots\end{cases},
\end{equation}
where $Q_k(t)$ represents the virtual control queue length for the $k$th service at the $t$th time slot. Based on this definition, we present the following theorem.

\begin{theorem}
The long-term average task processing cost constraint (\ref{III-a1}) is satisfied when the virtual queue remains stable, which occurs when $\mathop {\emph{lim} }\limits_{T\to\infty}\mathbb{E}[Q_{k}(T)]/T=0$.
\end{theorem}
\begin{proof}
The detailed proof is given in Appendix B of the supplemental material.
\end{proof}

\subsubsection{Maintain Queue Stability}
\label{SubSubSection: Maintain Queue Stability}

\par To stabilize the virtual queue, the Lyapunov function is first defined as:
\begin{equation}
\label{IV-a2}
L(Q_k(t))=\frac12Q_k(t)^2.
\end{equation}

\par The above equation represents a scalar measure of virtual queue congestion for task processing cost, where a smaller $L(Q_k(t))$ indicates less virtual queue backlog. Thus, maintaining the Lyapunov function at a bounded level through a policy implies that the virtual queue is stable. To this end, we introduce the Lyapunov drift, defined as:
\begin{equation}
\label{IV-a3}
\Delta(Q_k(t))=\mathbb{E}[L(Q_k(t+1))-L(Q_k(t))| Q_k(t)].
\end{equation}
$\Delta(Q_k(t))$ represents the increase in virtual queue backlog from time slot $t$ to $t+1$, which is determined by the task scheduling decision.

\subsubsection{Drift-plus-penalty Algorithm}
\label{SubSubSection: Drift-plus-penalty Algorithm}

\par After introducing the virtual queue, the original problem can be transformed into optimizing a new objective function, while ensuring that constraint in the current time slot is maintained. A practical approach is to minimize the Lyapunov drift plus penalty function, i.e., $\min \Delta(Q_k(t)) + VT_k(t)$ in each time slot. This incorporates queue stability into task response latency performance, with the weight $V$ balancing the trade-off between the two. Furthermore, the following theorem provides performance guarantees for this drift-plus-penalty function:
\begin{theorem}
For $\Delta(Q_k(t))$, generated by applying any task scheduling policy in each time slot, the following equation holds:
\begin{equation}
\label{IV-a3}
\Delta(Q_k(t))+VT_k(t)\leq B+VT_k(t)+Q_k(t)(E_k(t)-E_k^{avg}),
\end{equation}
where $ B=\frac{1}{2}E_{k}^{\max2}$ is a constant across all time slots, and $E_k^{max}=\max E_k(t)-E_k^{avg}$.
\end{theorem}
\begin{proof}
The detailed proof is given in Appendix C of the supplemental material.
\end{proof}

\par According to Theorem 3, the original problem can be solved by minimizing the right-hand side of (\ref{IV-a3}) at each time slot, transforming the problem into $\mathcal{P}_2$. 

\begin{equation}
\begin{aligned}\label{P2}
\mathcal{P}_2:&\mathop {\emph{min} }\limits_{\mathcal{N}_k(t)} B+VT_k(t)+Q_k(t)(E_k(t)-E_k^{avg}) \\
\mbox{s.t.}\;&(a): (\ref{II-a1}),(\ref{II-a2}),(\ref{II-b4}),(\ref{II-c4}),(14b).
\end{aligned}
\end{equation}

\par This approach eliminates the need for information from future time slots to calculate $\Delta(\Theta_k(t))$. However, it does not yield the optimal solution to the original problem. Later, we will analyze the gap between the solution from this method and the original optimal solution. Algorithm 1 provides the specific form of the drift-plus-penalty algorithm.

\begin{figure}[!t]
	\removelatexerror
	\begin{algorithm}[H]
		\raggedright
		\caption{Drift-plus-penalty Algorithm}
		\label{Algorithm 1: Drift-plus-penalty Algorithm}
		\LinesNumbered
				At the beginning of the $t$th time slot, obtain the distribution of request arrivals, resource allocations, and the value of the virtual queue for service $k$; \\
				Determine the optimal task scheduling decision $\mathcal{N}_k(t)^*$ by solving problem $\mathcal{P}_2$; \\
                Perform task scheduling based on $\mathcal{N}_k(t)^*$; \\
				Calculate $Q_k(t+1)$ according to Eq.(\ref{IV-a1}); \\
                $t\leftarrow t+1$; \\
                \textbf{goto} line 1;\\
	\end{algorithm}
\end{figure}

\subsection{A Heuristic-Based Hierarchical Optimization Approach}
\label{SubSection: A Heuristic-Based Hierarchical Optimization Approach}

\par In this subsection, we derive the optimal solution for problem $\mathcal{P}_2$ in each time slot. As stated in Theorem 4, it is not feasible to find an optimal solution to problem $\mathcal{P}_2$ within polynomial time \cite{31}.

\begin{theorem}
The problem $\mathcal{P}_2$ is NP-hard.
\end{theorem}
\vspace{-0.3cm}
\begin{proof}
The detailed proof is given in Appendix D of the supplemental material.
\end{proof}

\par To efficiently solve problem $\mathcal{P}_2$, Theorem 1 allows us to classify all factories/regions into three categories under the optimal task scheduling decision: source nodes, sink nodes, and isolated nodes. Once the categories of factories/regions are determined, we only need to decide the number of task requests dispatched from each source node to each sink node, thereby reducing the search space under the constraints of Theorem 1. Next, we first determine the optimal task scheduling decision for a given factory/region category assignment and then solve for the optimal category assignment scheme.

\subsubsection{Optimal Task Scheduling for a Given Factory/Region Category Assignment}
\label{SubSubSection: Optimal Task Scheduling for a Given Factory/Region Category Assignment}
Given a factory/region category assignment solution, represented by $\mathcal{O}$, $\mathcal{P}$, and $\mathcal{Q}$, where $\mathcal{O}$ indicates isolated nodes, $\mathcal{P}$ represents source nodes, and $\mathcal{Q}$ denotes sink nodes, the problem then reduces to determining the optimal number of task requests forwarded from each source node to each sink node. This solution should satisfy all constraints while minimizing the objective function $\mathcal{P}_2$. At this stage, $\mathcal{P}_2$ becomes
\begin{equation}
\begin{aligned}\label{P3}
\mathcal{P}_3:&\mathop {\emph{min} }\limits_{\{N_{p,q}^k(t)|p\in\mathcal{P},q\in\mathcal{Q}\}}B+VT_k(t)+Q_k(t)\Big(E_k(t)-E_k^{a\nu g}\Big) \\
\mbox{s.t.}\;&(a): \sum_{q\in\mathcal{Q}}N_{p,q}^k(t)\leq N_p^k(t),\forall p\in\mathcal{P},F_p^k(t)>0 \\
&(b): \sum_{q\in\mathcal{Q}}N_{p,q}^k(t)=N_p^k(t),\forall p\in\mathcal{P},F_p^k(t)=0 \\
&(c): N_{p,q}^k(t) \in \mathbb{N}, \forall p\in\mathcal{P},  \forall q\in\mathcal{Q}  \\
&(d): (\ref{II-b4}),(\ref{II-c4})
\end{aligned}
\end{equation}

\par $\mathcal{P}_3$ is an integer nonlinear programming (INLP) problem, for which we can apply the harmony search (HS) algorithm. As an intelligent optimization method, HS simulates a musician's improvisation process to find the optimal strategy for achieving harmony \cite{32}. To apply the HS algorithm to the constrained optimization problem $\mathcal{P}_3$, we introduce a penalty function, transforming $\mathcal{P}_3$ into $\mathcal{P}_4$:
\begin{equation}
\begin{aligned}\label{P4}
\mathcal{P}_4:&\mathop {\emph{min} }\limits_{\{N_{p,q}^k(t)|p\in\mathcal{P},q\in\mathcal{Q}\}}f(\{N_{p,q}^k(t)|p\in\mathcal{P},q\in\mathcal{Q}\}) \\
\mbox{s.t.}\;&(a): (\ref{II-b4}),(\ref{II-c4}),(20c)
\end{aligned}
\end{equation}
where $f\big(\big\{N_{p,q}^k(t)\big|p\in\mathcal{P},q\in\mathcal{Q}\big\}\big)= B+VT_k(t)+Q_k(t)(E_k(t)-E_k^{\alpha\nu g})+\frac{1}{\mu}(\sum_{p\in\mathcal{P}}\mathbb{I}(F_q^k(t)>0)max(0,\sum_{q\in\mathcal{Q}}N_{p,q}^k(t)-N_p^k(t))+\sum_{p\in P}\mathbb{I}(F_q^k(t)=0)max(0,|\sum_{q\in\mathcal{Q}}N_{p,q}^k(t)-N_p^k(t)|))$ and $\mu>0$. $\mathbb{I}(\cdot)$ is the indicator function and equals 1 (resp., 0) if the condition is true (resp., false). As $\mu$ approaches 0, the optimal solutions of $\mathcal{P}_4$ and $\mathcal{P}_3$ become increasingly similar.

\par The HS algorithm mainly involves the following basic parameters:
\begin{itemize}
\item{\emph{Harmonic memory size (HMS)}: The harmonic memory (HM) is modeled as a $HMS \times |\mathcal{P}||\mathcal{Q}|$ matrix, where each row is a solution to the problem $\mathcal{P}_4$;}
\item{\emph{Number of improvisations (NI)}: The maximum number of iterations to obtain the optimal harmony;}
\item{\emph{Harmonic memory consideration rate (HMCR)}: The probability of taking out a harmony from harmonic memory;}
\item{\emph{Pitch adjusting rate (PAR)}: The probability of adjusting pitch;}
\item{\emph{Step width (SW)}: The adjustment range of pitch;}
\end{itemize}

\par For illustration, let $ sol^{j}=\{N_{p,q}^{k,j}(t)\mid p\in\mathcal{P},q\in\mathcal{Q}\}$ represent the $j$th row of the HM, where $ j\in\{1,2\ldots, HMS\}$. $sol^B$ and $sol^W$ denote the best and worst solutions in the HM, while $UB(i)$ and $LB(i)$ represent the upper and lower bounds of the $i$th element, and $ sol^{j}(i)$ represents the $i$th element of $ sol^{j}$ ($i\in\{1, 2, \cdots, |\mathcal{P}||\mathcal{Q}|\}$). Algorithm 2 summarizes the detailed algorithmic procedure.

\begin{figure}[!t]
	\removelatexerror
	\begin{algorithm}[H]
		\raggedright
		\caption{The Implementation of HS Algorithm}
		\label{Algorithm 2: The Implementation of HS Algorithm}
		\LinesNumbered
		\KwIn{Set parameters: $HMS$, $NI$, $SW$, $HMCR$, $PAR$, $UB(i)$, $LB(i)$}
            \KwOut{The optimal solution of $\mathcal{P}_4$}
				Initialize the harmonic memory based on each element's upper and lower bounds. Substitute each initialized solution into $\mathcal{P}_4$ to calculate the objective function value $f(\cdot)$ and derive $sol^B$ and $sol^W$; \\
				\For{$Itr = 1$ \KwTo $NI$}{
					Initialize the new solution $sol^{new}$ as an all-zero vector of length $|\mathcal{P}||\mathcal{Q}|$; \\
                    \For{$i = 1$ \KwTo $|\mathcal{P}||\mathcal{Q}|$}{
                        \If{$rand() < HMCR$}{
                            Randomly select a row from the harmonic memory and assign its $i$th element to $sol^{new}(i)$; \\
                        }\Else{
                             $sol^{new}(i)=random.uniform(LB(i),UB(i))$; \\
                        }
                        \If{$rand() < PAR$}{
                            $sol^{new}(i)+=random.uniform(-1,1)\times SW$; \\
                            $sol^{new}(i)=min\left(max\left(sol^{new}(i),LB(i)\right),UB(i)\right)$; \\
                        }
                    }
        				 \If{$f\left(sol^{new}\right)<f\left(sol^{W}\right)$}{
                            Remove $sol^W$ from harmony memory and add $sol^{new}$; \\
                            Update $sol^B$ and $sol^W$; \\
                        }
				}
        \Return $sol^B$
	\end{algorithm}
\end{figure}

\par Notably, the optimal solution for $\mathcal{P}_4$ may not satisfy the constraints of $\mathcal{P}_3$. Therefore, after Algorithm 2 provides an initial solution, adjustments are necessary to ensure it meets $\mathcal{P}_3$'s constraints. This adjustment process aims to make the final solution compliant with the constraints while maintaining proximity to the initial result. Various adjustment methods are available, though details are omitted here.

\subsubsection{Optimal category assignment scheme}
\label{SubSubSection: Optimal category assignment scheme}

\par We can now derive the optimal task scheduling policy when the factory/region category solution is given. The next step is to determine the optimal category assignment solution. Based on the conditions outlined in Theorem 1, the optimal task scheduling decision allows for only three types of vertex categories. Therefore, we develop an enhanced discrete particle swarm algorithm to determine the optimal vertex category assignment. In this research, a swarm of $N$ particles moves through an $M$-dimensional search space at a defined speed. Each particle adjusts its position (i.e., vertex category assignment solution) based on its historical best point and the global best point of all particles in the swarm. The $i$th particle of the swarm ($i = 1, 2, 3,\cdots, N$) consists of the following three vectors:
\begin{itemize}
\item{\emph{Current position}: $v_{i}=(v_{i1},v_{i2},\ldots v_{iM})$;}
\item{\emph{Historical optimal position}: $p_{i}=(p_{i1},p_{i2},\ldots p_{iM})$;}
\item{\emph{Velocity}: $v_{i}=(v_{i1},v_{i2},\ldots v_{iM})$.}
\end{itemize}

\par To improve the global search efficiency and accelerate the process, we impose constraints on each particle, requiring them to meet the conditions outlined in Corollary 1.
\begin{corollary}
The particle corresponding to the optimal category assignment should satisfy the following conditions.
\begin{enumerate}
\item{Each element in a particle's position vector can only take values from the set $\mathcal{S}=\{0,1,2\}$, representing the corresponding vertex as a source, sink, or isolated node, respectively.}
\item{The elements in a particle’s position vector corresponding to the set $\{m|F_m^k(t)=0\}$ can only take the value 0.}
\item{If a position vector contains an element with the value 0, it must also contain an element with the value 1.}
\end{enumerate}
\end{corollary}
\par The three conditions above are derived from Theorem 1.1), 2), and 4), respectively. Additionally, for each particle, the velocity in its $m$th dimension changes according to the following equation:
\begin{equation}
\begin{aligned}
\label{IV-b1}
v_{im}=w_mv_{im}&+c_1\cdot\mathrm{rand}()\cdot(p_{im}-x_{im})\\
&+c_2\cdot\mathrm{rand}()\cdot(p_{gm}-x_{im}),
\end{aligned}
\end{equation}
where $p_{g}=(p_{g1},p_{g2},\ldots p_{gM})$ represents the global optimal position of the swarm. The parameter $w_m$ is the inertia weight factor, while $c_1$ and $c_2$ are acceleration constants. These three non-negative parameters balance the trade-off between exploration and exploitation, with $w_m + c_1 + c_2 = 1$.

\par The equation governing the variation in the $m$th dimension of each particle's position is shown below:
\begin{equation}
\label{IV-b2}
X_{im}=\begin{cases}\text{random }(S\setminus\{X_{im}\}),&\text{if} \;rand()<\frac{1}{1+e^{-v_{im}}}\\X_{im},&\text{else}\end{cases},
\end{equation}
where $\text{random }(S\setminus\{X_{im}\})$ indicates a random selection of an element from the set $ S\setminus\{X_{im}\}$.

\begin{figure}[!t]
	\removelatexerror
	\begin{algorithm}[H]
		\raggedright
		\caption{Category Assignment Based on an Enhanced Discrete Particle Swarm Algorithm}
		\label{Algorithm 3: Category Assignment Based on an Enhanced Discrete Particle Warm Algorithm}
		\LinesNumbered
		\KwIn{Particle swarm size $N$, number of factories $M$, inertia weight $w_m$, acceleration constants $c_1, c_2$, maximum number of iterations $I_{max}$}
            \KwOut{Optimal category assignment solution}
				Randomly initialize the positions and velocities of $N$ particles within the $M$-dimensional problem space. Each particle must satisfy the conditions in Corollary 1, and start with an individual optimal fitness $ pbest_{i}=0$ and a global optimal fitness $gbest=0$; \\
				\For{$Itr = 1$ \KwTo $I_{max}$}{
                        \For{$i = 1$ \KwTo $N$}{
                             For the $i$th particle, the category assignment solution is determined from its position vector. Algorithm 2 is used to derive the optimal scheduling decision for this category assignment solution, yielding the objective function value $H$ of $\mathcal{P}_3$. The particle's fitness is then calculated as $1/H$; \\
                             \If{$1/H>pbest_{i}$}{
                                Update the particle's historical best position $p_i$ to the current position and $pbest_i = 1/H$; \\
                            }
                            \If{$1/H>gbest$}{
                                Update $p_g$ to the current position and $gbest= 1/H$; \\
                            }
                            Update $i$th particle's velocity and position according to equations (\ref{IV-b1}) and (\ref{IV-b2}); \\
                            If the modified particle does not meet the conditions outlined in Corollary 1, revert to the previous step and re-execute the process until it does; \\
            }
        }
        \Return $p_g$
	\end{algorithm}
\end{figure}

\par Algorithm 3 outlines the steps for determining the optimal category assignment solution.

\par
\begin{figure*}[htbp]
	\centering	\includegraphics[width=\linewidth,scale=1.00]{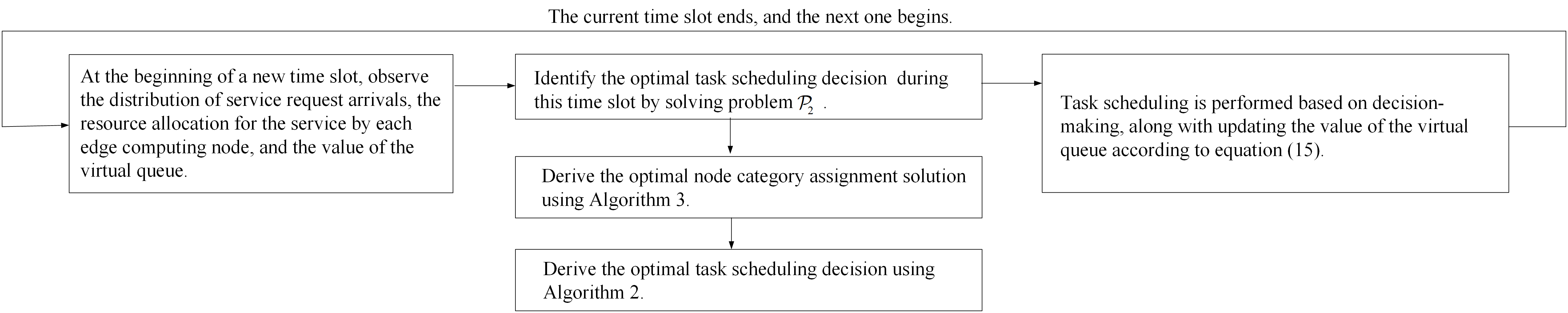}
	\caption{Online task scheduling framework.}
	\label{Figure:Algorithm Framework2}
\end{figure*}

 At this point, we have outlined all foundational algorithms within the online task scheduling framework. The overall flow of the proposed framework is illustrated in Fig. \ref{Figure:Algorithm Framework2}.

\subsection{An Imitation Learning-Based Low-Complexity Approach}
\label{SubSection: An Imitation Learning-Based Low-Complexity Approach}

\par Although the heuristic-based hierarchical optimization approach reduces the search space for the optimal solution, determining the optimal factory/region category assignment scheme still increases execution latency, particularly in large systems (e.g., multiple factories). Section \ref{Section:Performance Evaluation} compares algorithm execution times and demonstrates that the search for the optimal category assignment scheme significantly contributes to total execution delay. Considering the requirement for low-complexity algorithms in IIoT scenarios, this section proposes an imitation learning-based approach.

\par Imitation learning is a reinforcement learning concept that enables an intelligent agent to achieve expert-level performance on a given task by mimicking an expert’s behavior. The core idea of imitation learning is to enable an intelligent agent to learn an expert’s decision-making patterns without requiring an explicit reward function \cite{+16in1stMarjor}. Imitation learning primarily comprises two methods: behavior cloning (BC) and inverse reinforcement learning (IRL). Behavior cloning is the most straightforward approach, treating imitation learning as a supervised learning problem in which an intelligent agent is trained on an expert’s state-action data pairs to make similar decisions in the same state \cite{+17in1stMarjor}. Inspired by this, we can model the decisions of the heuristic-based hierarchical optimization approach as expert behaviors and represent the intelligent agent using a deep neural network. 

\par First, we need to specify the state $\mathcal{S}_t$ and action $\mathcal{A}_t$. Based on Section \ref{Section:Problem Formulation and Analysis}, we define the system state graph as the state, merging its two global attributes into $Q_k(t)$ according to Equation (\ref{IV-a1}). The action is a $|\mathcal{M}|$-dimensional vector, where each element takes a value from $\{0,1,2\}$, representing a vertex as a source, sink, or isolated node, respectively. We do not directly use the task scheduling decision graph as the action due to its complex constraints (as shown in Theorem 1), which require more data and incur higher training costs to achieve acceptable accuracy. Experimental results indicate that the key factor affecting the algorithm’s running time is determining the optimal category assignment scheme. Therefore, we train the intelligent agent to learn this determination and subsequently solve the corresponding task-scheduling decision using Algorithm 2.

\par Specifically, the scheme follows these steps:

\begin{itemize}
\item{\emph{Step 1}: During the specified period, collect the expert’s state-action pairs, denoted as $\{\overline{\mathcal{S}}_t, \overline{\mathcal{A}}_t| t\in \mathcal{I}\}$, where $\overline{\mathcal{A}}_t$ represents the category assignment decision made by the heuristic-based hierarchical optimization approach.}
\item{\emph{Step 2}: Construct a deep neural network as an intelligent agent, consisting of multiple graph convolutional layers and $M$ multilayer perceptrons (MLPs), with the system state graph as input. First, the message-passing mechanism of the graph convolutional layers is used to obtain the embedding representation of each vertex. The embedding representation of the $m$th vertex is then fed into the $m$th MLP, whose output dimension is 3, corresponding to the number of vertex categories. Finally, a softmax activation function is applied to obtain the probability of each vertex belonging to different categories.}
\item{\emph{Step 3}:  A stochastic gradient descent optimization algorithm (e.g., Adam) is employed to train the intelligent agent constructed in Step 2 using the expert dataset $\{\overline{\mathcal{S}}_t, \overline{\mathcal{A}}_t| t\in \mathcal{I}\}$. The training process minimizes a cross-entropy loss function: $\mathcal{L}=-\sum_{m\in\mathcal{M}}\log(\hat{y}_m)$, where $\hat{y}_m$ denotes the probability predicted by the agent that vertex $m$ belongs to its true label category.}
\item{\emph{Step 4}: Save the trained model parameters, deploy the model to the central controller, and load the trained parameters.}
\item{\emph{Step 5}: During deployment, the intelligent agent predicts the category $\mathcal{A}_t$ of all nodes based on the current system state graph $\mathcal{S}_t$. Given the predicted category assignment scheme, Algorithm 2 is executed to determine the task scheduling decision.}
\end{itemize}

\section{Theoretical Analysis}
\label{Section:Theoretical Analysis}

\subsection{Computational Complexity}
\label{SubSection:Computational Complexity}

\subsubsection{Heuristic-Based Hierarchical Optimization Approach}
\label{SubSubSection: Heuristic Hierarchical Optimization Framework}
\par As illustrated in Fig. \ref{Figure:Algorithm Framework2}, the task scheduling decision generation process in each time slot involves a two-stage heuristic algorithm. The outer layer addresses the optimal category assignment using an enhanced particle swarm algorithm, while the inner layer determines the optimal scheduling decision for a given category assignment via the harmonic search algorithm. In Algorithm 3, the outer layer iterates up to $I_{max}$ times, traversing $O(N)$ node category assignment solutions in each iteration, where adjusting each category assignment solution requires traversing $M$ dimensions. Additionally, each category assignment solution requires the execution of Algorithm 2 to compute its corresponding fitness. 

\par In Algorithm 2, up to $NI$ iterations are performed. During each iteration, a harmony is generated, requiring $O(M^2)$ traversals. Therefore, the computational complexity of Algorithm 2 is $O(NI\times M^2)$.

\par Based on the above analyses, the computational complexity of generating task scheduling decisions in each time slot is $O(I_{max}\times N\times (M+NI \times M^2))$, where $I_{max}$, $NI$, and $N$ are algorithm hyperparameters.

\subsubsection{Imitation Learning-Based Scheme}
\label{SubSubSection: Imitation Learning Scheme}
\par In this scheme, the deep neural network first predicts the category assignment scheme, and then Algorithm 2 generates the task scheduling decision based on the given category assignment scheme. In the deep neural network, the node embedding representation is obtained using graph convolutional layers with a complexity of $O(H\times M^{2})$, where $H$ is the hidden layer dimension. Each node category prediction is performed by a two-layer MLP with a complexity of $O(H^2)$.

\par Considering the computational complexity of Algorithm 2, the overall complexity of the scheme can be derived as $O(H\times M^{2}+H^2+NI\times M^{2})$. This demonstrates that the proposed scheme lowers the overall computational complexity by reducing the number of executions of Algorithm 2.

\subsection{Convergence Performance}
\label{SubSection:Convergence Performance}
\par Using the proposed algorithms, the near-optimal solution to problem $\mathcal{P}_2$ can be obtained. The convergence of problem $\mathcal{P}_1$ is guaranteed under our online algorithm. Assuming problem $\mathcal{P}_1$ is a bounded function with upper and lower limits $T^{max}_k$ and $T^{min}_k$, respectively. We define $\tilde{T}_{k}(t)$ as the delay performance of the proposed algorithm in time slot $t$, and $T^{opt}_k$ as the lower bound of the time-averaged delay performance with global information. The following theorem provides an upper bound on the time-averaged delay performance and the task scheduling cost queue backlog.
\begin{theorem}
For any non-negative control parameter $V$, the long-term delay performance achieved by the online algorithm satisfies the following condition:
\begin{equation}
\label{V-b1}
\lim_{T\to\infty}\frac1T\sum_{t=0}^{T-1}\mathbb{E}\{\tilde{T}_k(t)\}\leq T^{opt}_k+\frac{B+D}V,
\end{equation}
where $D$ represents the upper bound of the performance gap between the solution of problem $\mathcal{P}_2$ obtained using our proposed algorithms and the optimal solution of problem $\mathcal{P}_1$ under global information in each time slot.
\end{theorem}
\begin{theorem}
Given $E_k^{avg} > 0$ and that the initial task processing cost virtual queue backlog is $0$, we have the following upper bound for the virtual queue backlog:
\begin{equation}
\label{V-b2}
\lim_{T\to\infty}\frac1T\sum_{t=0}^{T-1}\mathbb{E}\{Q(t)\}\leq\frac{B+V(T^{max}-T^{min})+D}\varepsilon,
\end{equation}
where $ \varepsilon>0$ is a bounded constant representing the difference between the time-averaged task processing cost under a specific control strategy and the long-term cost budget.
\end{theorem}

\par Due to space constraints, the proof process is not elaborated upon here; however, the detailed proof is analogous to that in \cite{33}. According to Theorem 4, the delay performance of the online algorithm gradually approaches the offline optimal solution as the controllable parameter $V$ increases. Additionally, as shown in Theorem 5, the upper bound of the virtual queue backlog for task processing cost is also determined by the parameter $V$. In summary, there is a performance-cost trade-off of $[O(1/V), O(V)]$ for the online algorithm. Therefore, we can adjust the parameter $V$ to achieve a balance between long-term delay performance and the operational cost.

\section{Performance Evaluation}
\label{Section:Performance Evaluation}

\subsection{Simulation Setting}
\label{SubSection:Simulation Setting}

\par In the simulation, we consider the scenario shown in Fig. 1, where the number of cells is set to 5. According to Section \ref{Section:System Model}, simulating a single service is sufficient; thus, we only consider service $k$. The computational resources allocated to this service are spatially non-uniform. The number of requests arriving at each AP is spatio-temporally non-uniform. Additionally, the transmission bandwidth allocated to service $k$ between any two factories is between 10 and 100 Gbps, the service request size is between 200 and 500 Kbits, the CPU cycles required to process each request range from 50 to 500 Kilocycles, and the long-term average task processing cost budget is set to 20 J per time slot. The main parameters of the simulation are summarized in Table \ref{tab:Main Simulation Parameters}, with all parameters using the values provided unless otherwise specified. All simulations were executed on a Pytorch 1.10.2 platform with an Intel Core i7-13650HX 4.9 GHz CPU and 16 GB of RAM.

\begin{table}[h]
\centering
\caption{Main Simulation Parameters}
\label{tab:Main Simulation Parameters}
\resizebox{0.9\columnwidth}{!}{%
\begin{tabular}{l|l}
\hline
Parameters                & Value            \\ \hline
The computational resources allocated \\ by ES $1$ to service $k$ $F_1^k(t)$                & {[}50, 60{]} GHz \\
The computational resources allocated \\ by ES $2$ to service $k$ $F_2^k(t)$                & {[}40, 50{]} GHz \\
The computational resources allocated \\ by ES $3$ to service $k$ $F_3^k(t)$                & {[}30, 40{]} GHz \\
The computational resources allocated \\ by ES $4$ to service $k$ $F_4^k(t)$                & {[}20, 30{]} GHz \\
The computational resources allocated \\ by ES $5$ to service $k$ $F_5^k(t)$                & {[}10, 20{]} GHz \\
The number of requests arriving at AP $1$ \\ at the start of a time slot for service $k$ $N_1^k(t)$                & {[}10, 20{]}     \\
The number of requests arriving at AP $2$ \\ at the start of a time slot for service $k$  $N_2^k(t)$                & {[}20, 30{]}     \\
The number of requests arriving at AP $3$ \\ at the start of a time slot for service $k$  $N_3^k(t)$                & {[}30, 40{]}     \\
The number of requests arriving at AP $4$ \\ at the start of a time slot for service $k$  $N_4^k(t)$                & {[}40, 50{]}     \\
The number of requests arriving at AP $5$ \\ at the start of a time slot for service $k$  $N_5^k(t)$                & {[}50, 60{]}     \\
 the
Cost of scheduling a single request for \\ service $k$ from
AP $m$ to ES $m^{\prime}$ $E_{m,m^{\prime}}^{k}(t)$ & {[}0.4, 2.4{]}   \\
$V$                       & $10^8$           \\
Total number of time slots $T$                       & 1000           \\
Computational energy efficiency factor $k_m$                       & $10^{-26}$           \\ \hline
\end{tabular}%
}
\end{table}

\par To evaluate the performance of the proposed schemes, we compare them against the following benchmark schemes.
\begin{enumerate}
\item{\emph{Myopic} \cite{34}: The Myopic scheme ignores the queue backlog and minimizes the average processing delay of all requests for service $k$ in each time slot by solving:
\begin{equation}
\begin{aligned}\label{P4}
\mathcal{P}_4:&\mathop {\emph{min} }\limits_{\mathcal{N}_k(t)}\mathop {\emph{lim} }\limits_{T\to\infty}\frac1T\sum_{t=1}^TT_k(t) \\
\mbox{s.t.}\;&(a): (\ref{II-a1}),(\ref{II-a2}),(\ref{II-b4}),(\ref{II-c4}), \\
&(b): E_k(t)\leq tE_k^{avg}-\sum_{l=1}^{t-1}E_k(l)
\end{aligned}
\end{equation}

Here, constraint (b) ensures that the long-term average task processing budget constraint of $\mathcal{P}_1$ is met by the $t$-th time slot, where $\{E_k(l) \mid l = 1, 2, \cdots, t-1\}$ represents the known past task processing cost up to the $t$-th time slot.

}
\item{\emph{K time-step-greedy algorithm (KG)} \cite{36}: Task scheduling aimed at minimizing latency is performed every $K$ time slots, without accounting for energy consumption constraints.}
\item{\emph{No scheduling (NS)}: Tasks are processed immediately upon arrival at the edge computing node, without any task scheduling.}
\item{\emph{Load-balanced scheduling (LBS)} \cite{37}: Tasks arriving at each edge node are scheduled proportionally to the computational resources allocated to the service by each edge computing node.}
\item{\emph{Standard heuristic algorithm (SH)}}: The transformed problem $\mathcal{P}_2$ is directly solved using a heuristic algorithm, such as the genetic algorithm \cite{15}. The optimization variable is defined as the number of tasks each AP forwards to each ES.
\item{\emph{Deep reinforcement learning algorithm (DRL)}}: The proximal policy optimization (PPO) algorithm \cite{ +18in1stMarjor } is employed, where the state is defined as the system state graph. The action is defined as the proportion of tasks each AP forwards to each ES. The reward function is given by 
\begin{equation}
\label{VI-a1}
R(t)=-(VT_k(t)+max\{(E_k(t)-E_k^{avg}),0\}).
\end{equation}
\end{enumerate}

\begin{figure*}[t]
\centering
\subfloat[]{\includegraphics[width=0.25\textwidth]{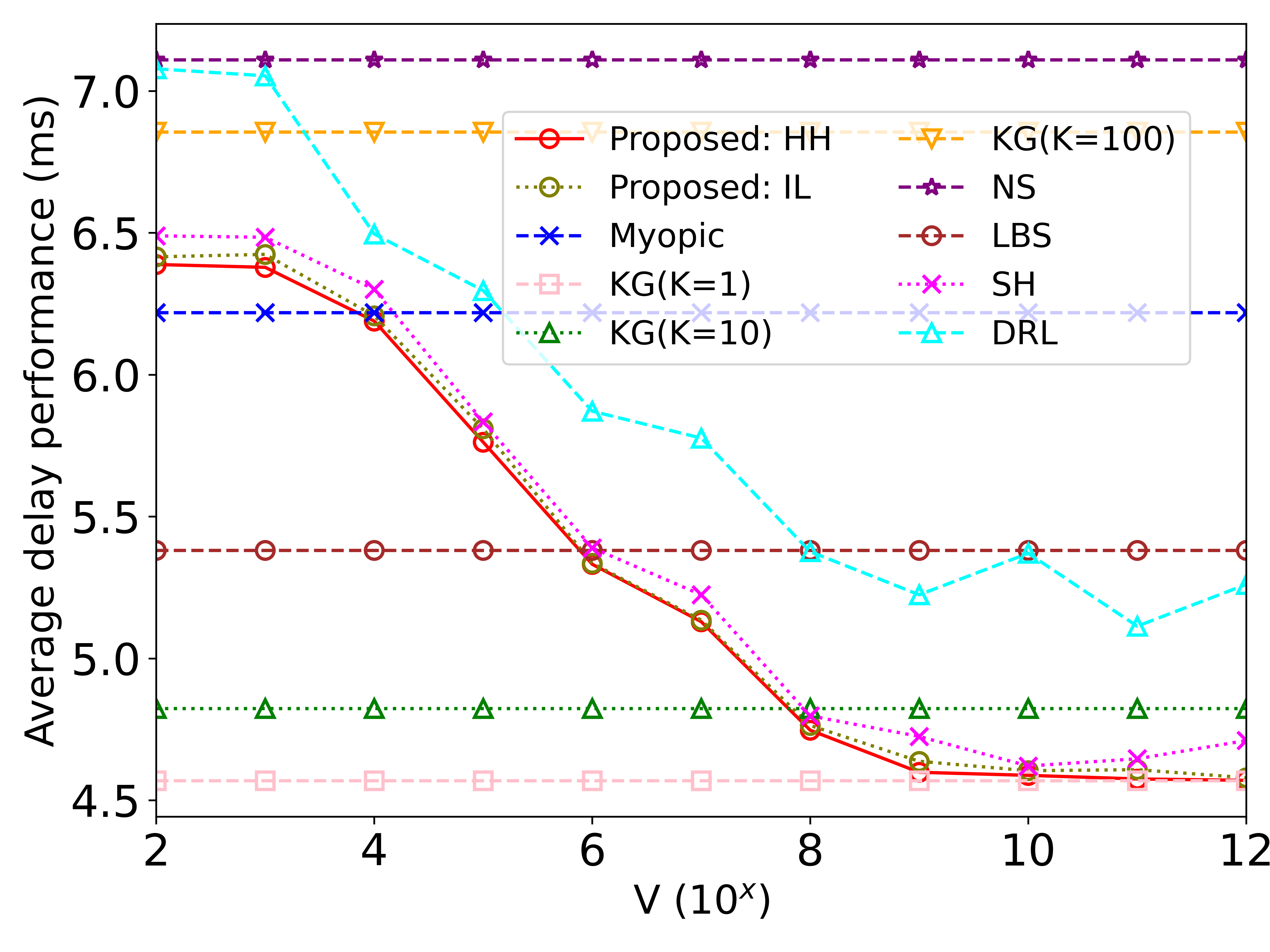}%
}
\subfloat[]{\includegraphics[width=0.25\textwidth]{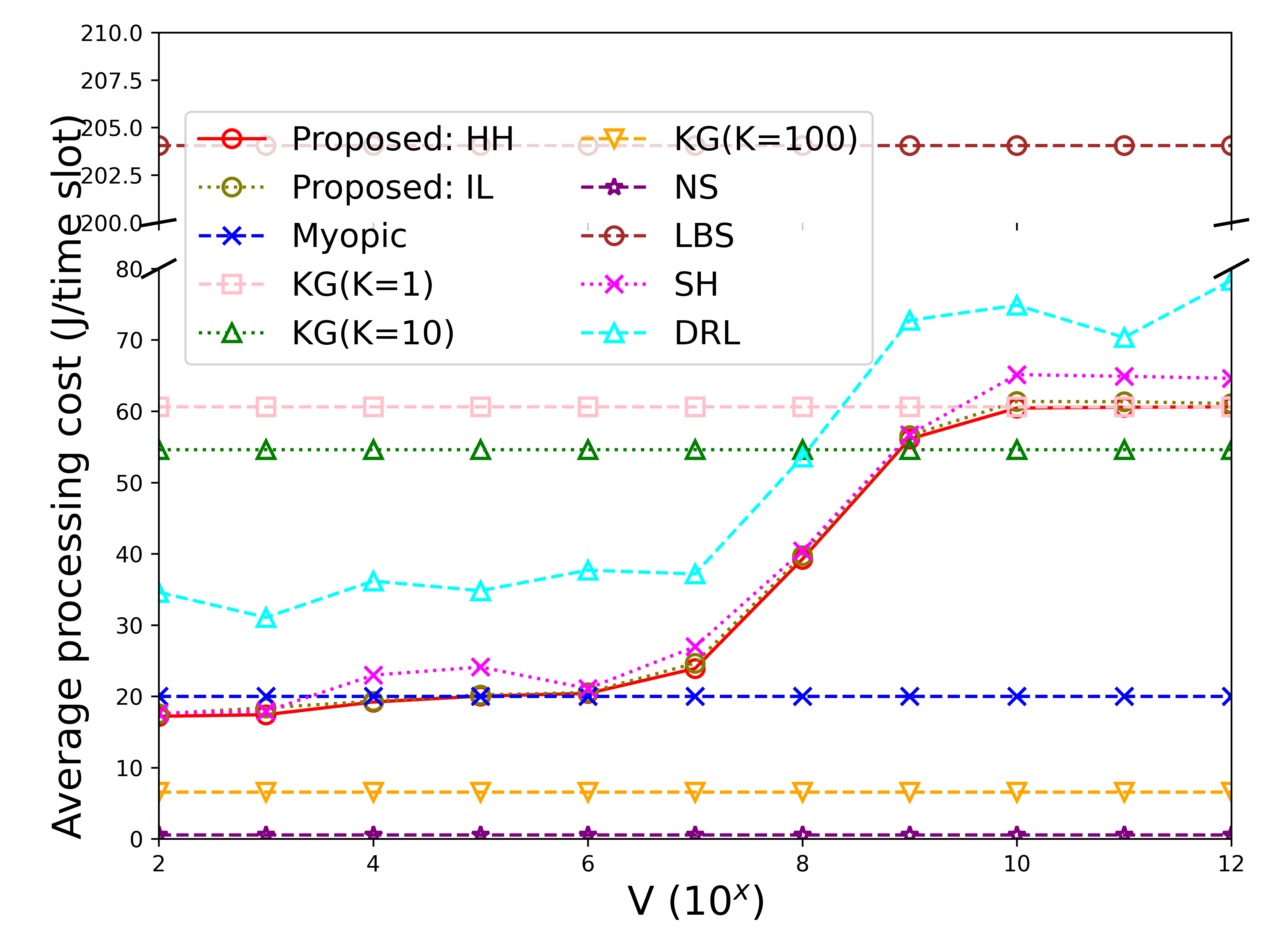}%
}
\subfloat[]{\includegraphics[width=0.25\textwidth]{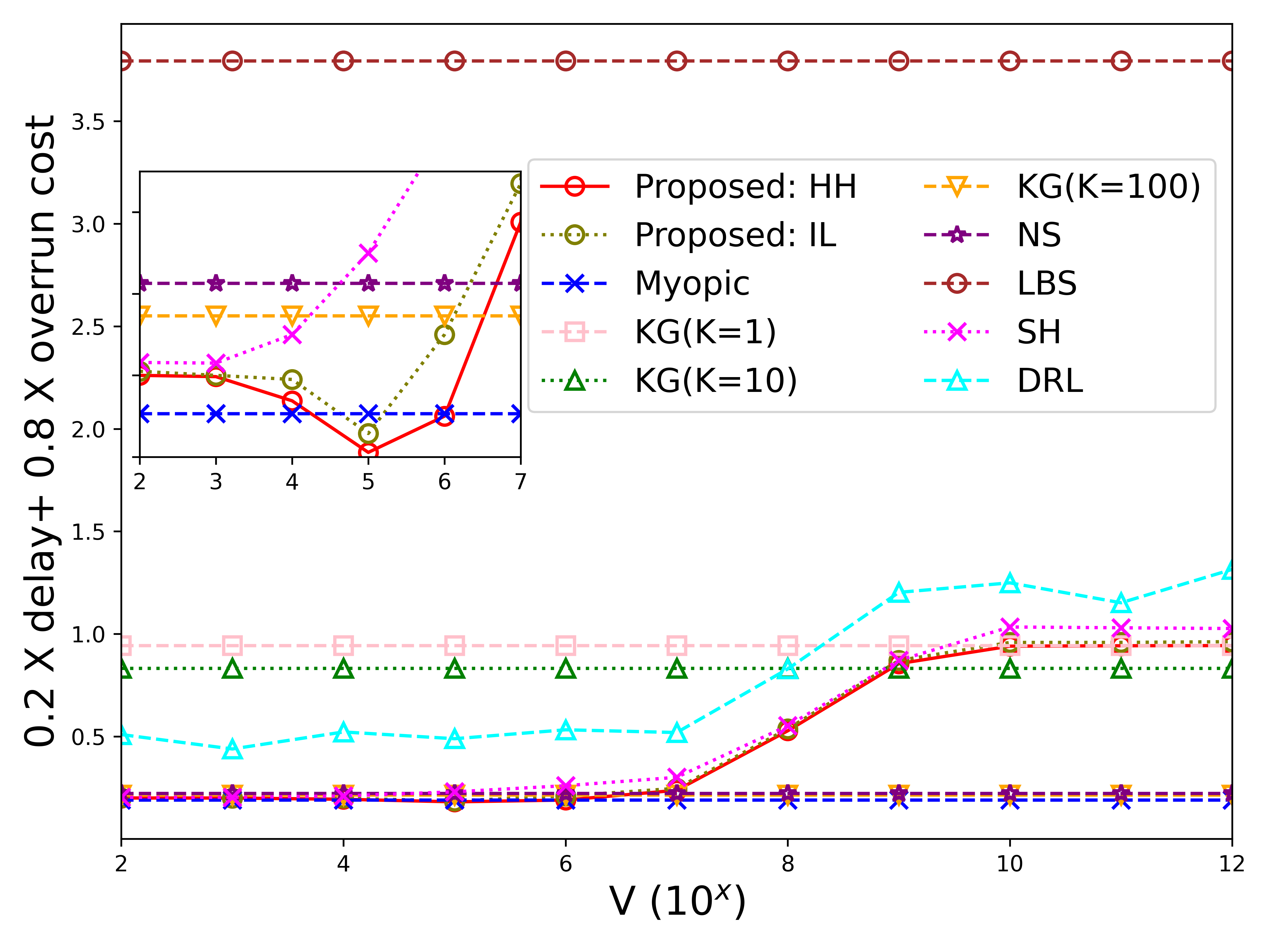}%
}
\subfloat[]{\includegraphics[width=0.24\textwidth]{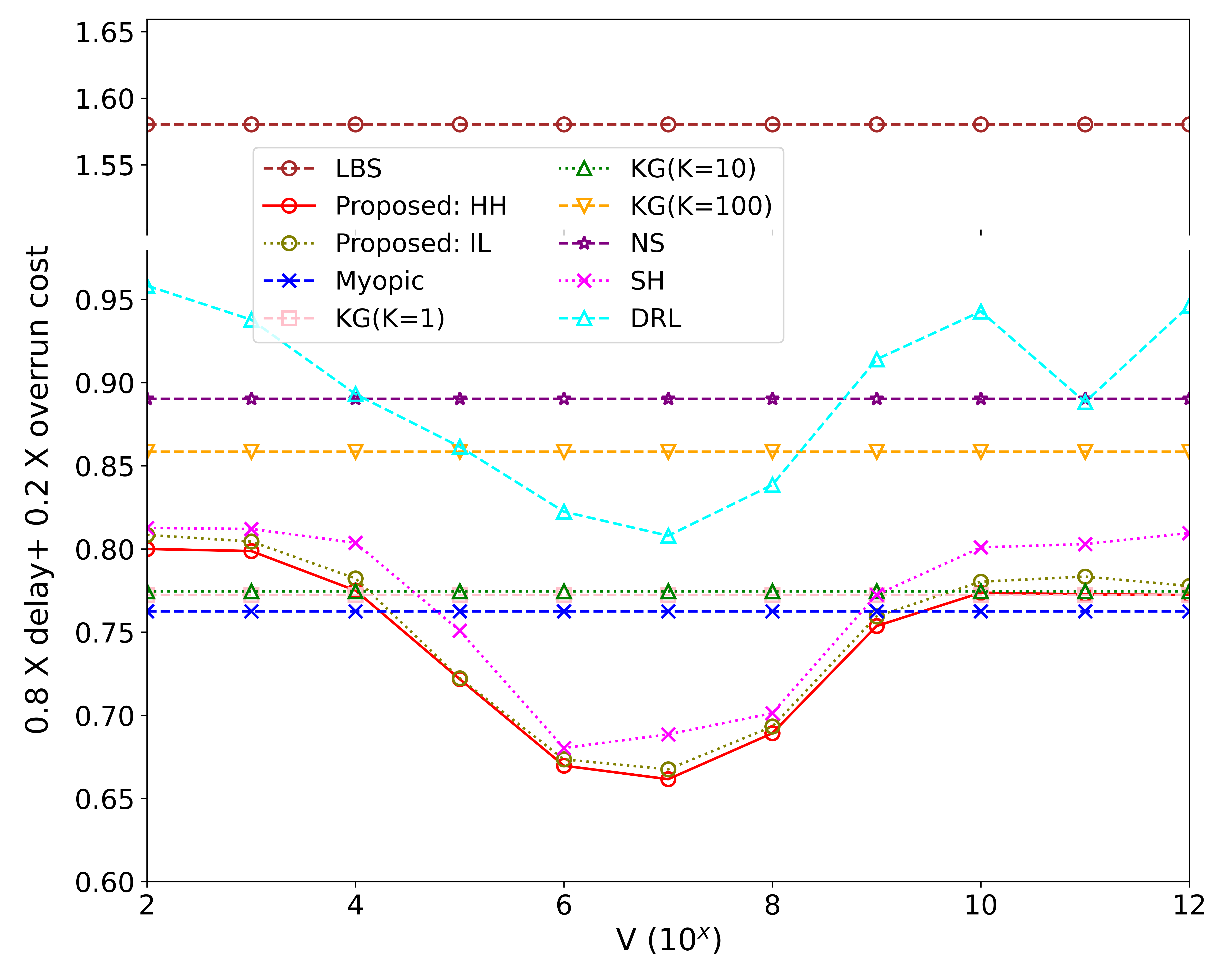}%
}

\caption{Performance comparison of different schemes: (a) Achievable delay performance under different schemes. (b) Achievable average task processing cost under different schemes. (c) Algorithm performance when the normalized delay and average overrun cost weights are 0.2 and 0.8, respectively. (d) Algorithm performance when the normalized delay and average overrun cost weights are 0.8 and 0.2, respectively. Note: The normalization of delay and average overrun cost is based on the maximum delay and highest overrun cost observed in the proposed scheme using the heuristic-based hierarchical optimization approach. The legends "Proposed: HH" and "Proposed: IL" represent the online task scheduling scheme based on heuristic hierarchical optimization and the online task scheduling scheme based on imitation learning, respectively.}
\label{Performance comparison}
\end{figure*}

\subsection{Simulation Results}
\label{SubSection:Simulation Results}

\par The detailed results of our simulation are provided in this section. We compare the effectiveness of the proposed schemes with several above-mentioned benchmark policies, and analyze the performance of the online task scheduling scheme based on heuristic hierarchical optimization under various algorithm configurations and environmental settings.

\subsubsection{Algorithm Performance Comparison}
\label{SubSubSection: Algorithm Performance Comparison}

\par Fig. \ref{Performance comparison} presents the performance of various schemes in terms of delay, average task processing cost, and the weighted sum of the normalized delay and average overrun cost. First, it can be seen that the performance of the two proposed schemes is nearly identical. Fig. \ref{Performance comparison}(a) and (b) show that as $V$ increases, the average task processing delay of the two proposed schemes decreases, whereas the long-term average task processing cost gradually rises. Additionally, the delay and average task processing cost of the two proposed schemes remain stable when $V$ is either very small or very large, corresponding to prioritizing task processing cost and delay, respectively. Both proposed schemes satisfy the long-term average task processing cost constraint when $V<10^6$. For $10^4<V<10^6$, the two proposed schemes achieve optimal delay performance under this constraint. However, LBS and KG ($K=1, 10$) algorithms fail to meet the long-term energy constraint despite providing better delay performance. As $K$ increases, the average task processing delay under the KG algorithm gradually rises, whereas the corresponding energy consumption decreases. Therefore, in the KG algorithm, $K$ has a similar effect to $V$ in the proposed schemes. Adjusting $K$ enables a trade-off between task processing delay and cost. However, the KG algorithm exhibits significant performance fluctuations over time slots, resulting in higher user experience variability. In contrast, the proposed schemes can provide a more stable user experience.

\par Fig. \ref{Performance comparison}(c) and (d) show the weighted sum of normalized delay and average overrun cost. In Fig. \ref{Performance comparison}(c), the objective function assigns a weight of 0.2 to delay and 0.8 to overrun cost, while in Fig. \ref{Performance comparison}(d), these weights are 0.8 and 0.2, respectively. As shown in Fig. \ref{Performance comparison}(c), the two proposed schemes achieve optimal performance for the given metric at $V = 10^5$. In Fig. \ref{Performance comparison}(d), the two proposed schemes achieve optimal performance for the given metric at $V = 10^7$. Therefore, the proposed schemes can be tuned to achieve optimal performance by adjusting the value of $V$ according to the weighting preferences for overrun cost and delay.

\par Additionally, the SH and DRL schemes consistently perform worse than the two proposed schemes, with the DRL scheme exhibiting significant volatility. This is because the SH scheme contains numerous redundant or invalid solutions in its search space, reducing optimization efficiency. In contrast, the DRL scheme suffers from low sample efficiency and an unstable training process, causing early-stage strategies to be random and volatile.
\subsubsection{Scalability}
\label{Scalability}

\begin{figure*}[t]
\centering
\subfloat[]{\includegraphics[width=0.31\textwidth]{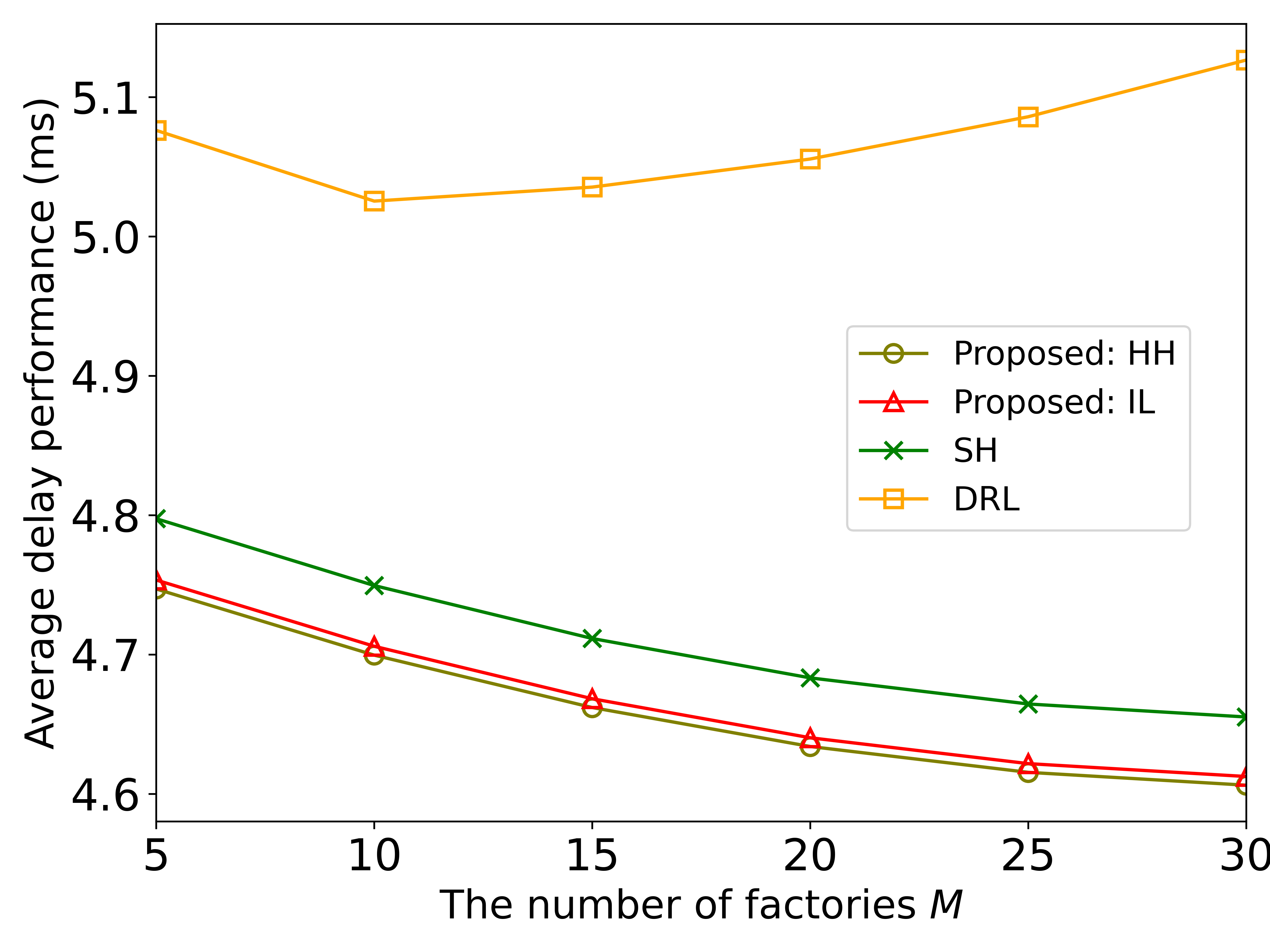}%
}
\subfloat[]{\includegraphics[width=0.31\textwidth]{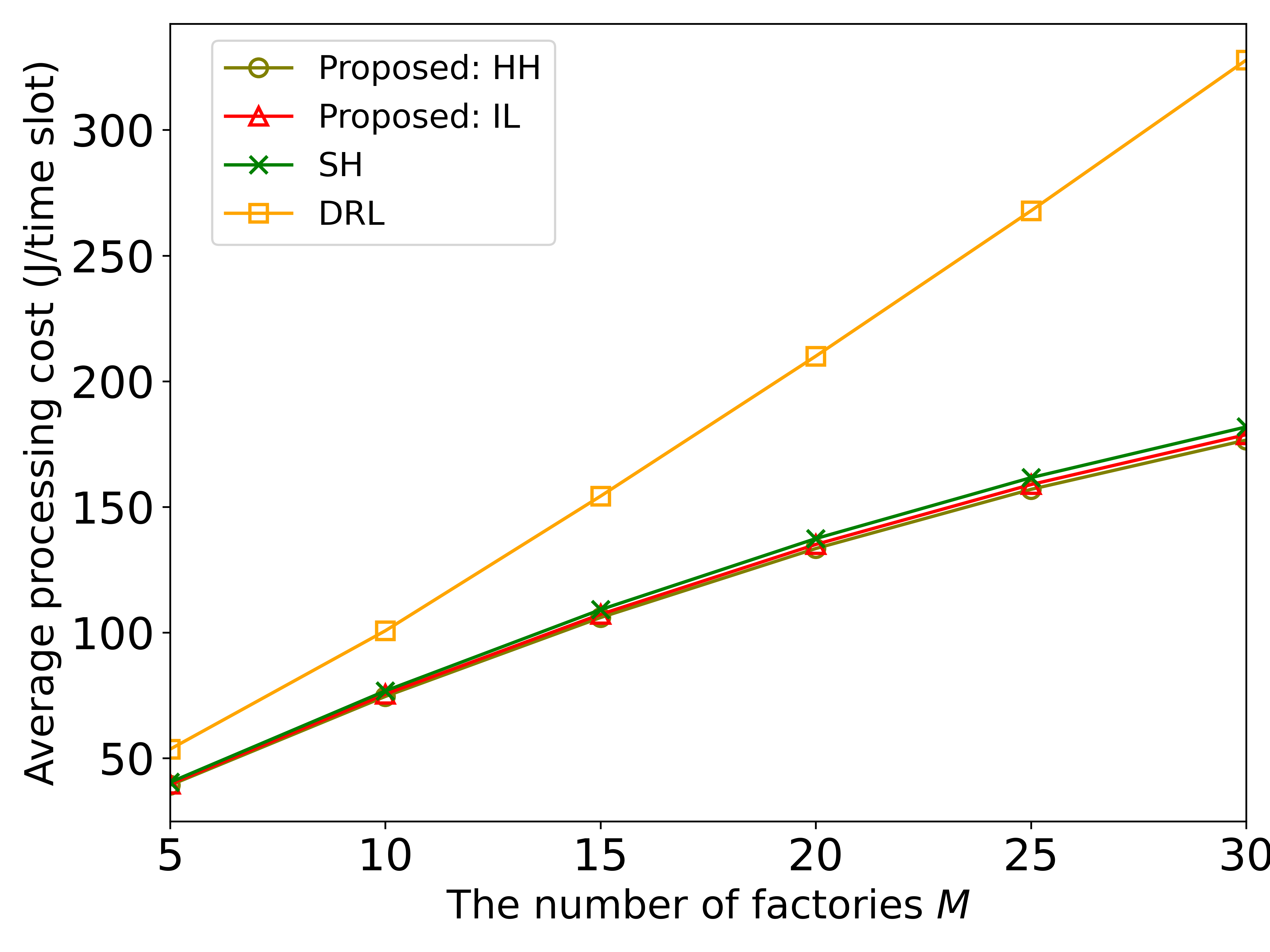}%
}
\subfloat[]{\includegraphics[width=0.31\textwidth]{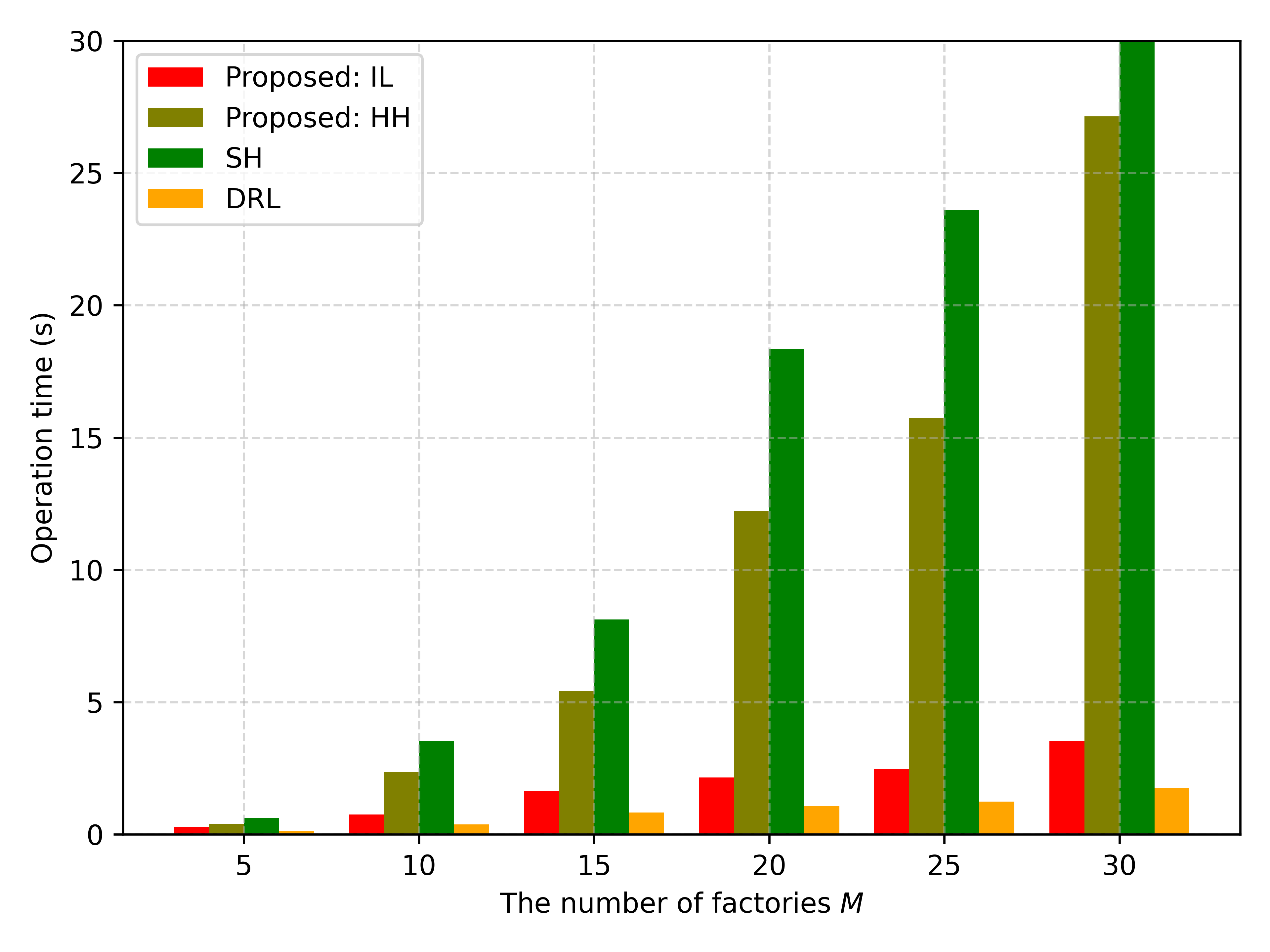}%
}

\caption{Scalability comparison of different schemes: (a) Achievable delay performance variation with the number of factories. (b) Average task processing cost variation with the number of factories. (c) Average algorithm running time variation with the number of factories. Note: In Fig. \ref {Scalability Fig} (a) and (b), the number of iterations for the SH algorithm is set to match the total iterations of the HH algorithm. In Fig. \ref {Scalability Fig} (c), the stopping condition for the SH algorithm is defined as reaching a target value as close as possible to that of the HH algorithm.}
\label{Scalability Fig}
\end{figure*}

\begin{figure*}[t]
\centering
\subfloat[]{\includegraphics[width=0.33\textwidth]{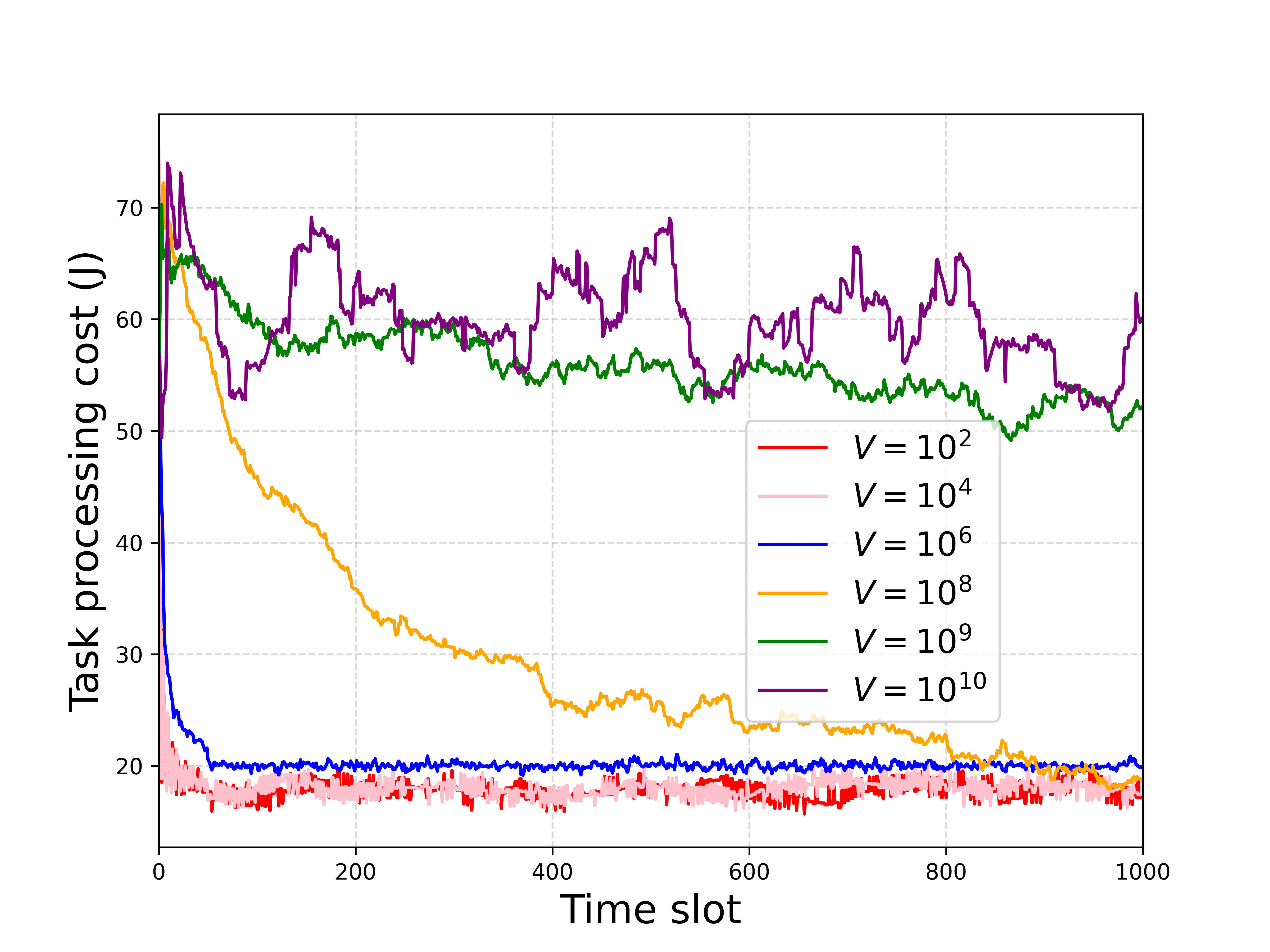}%
}
\subfloat[]{\includegraphics[width=0.33\textwidth]{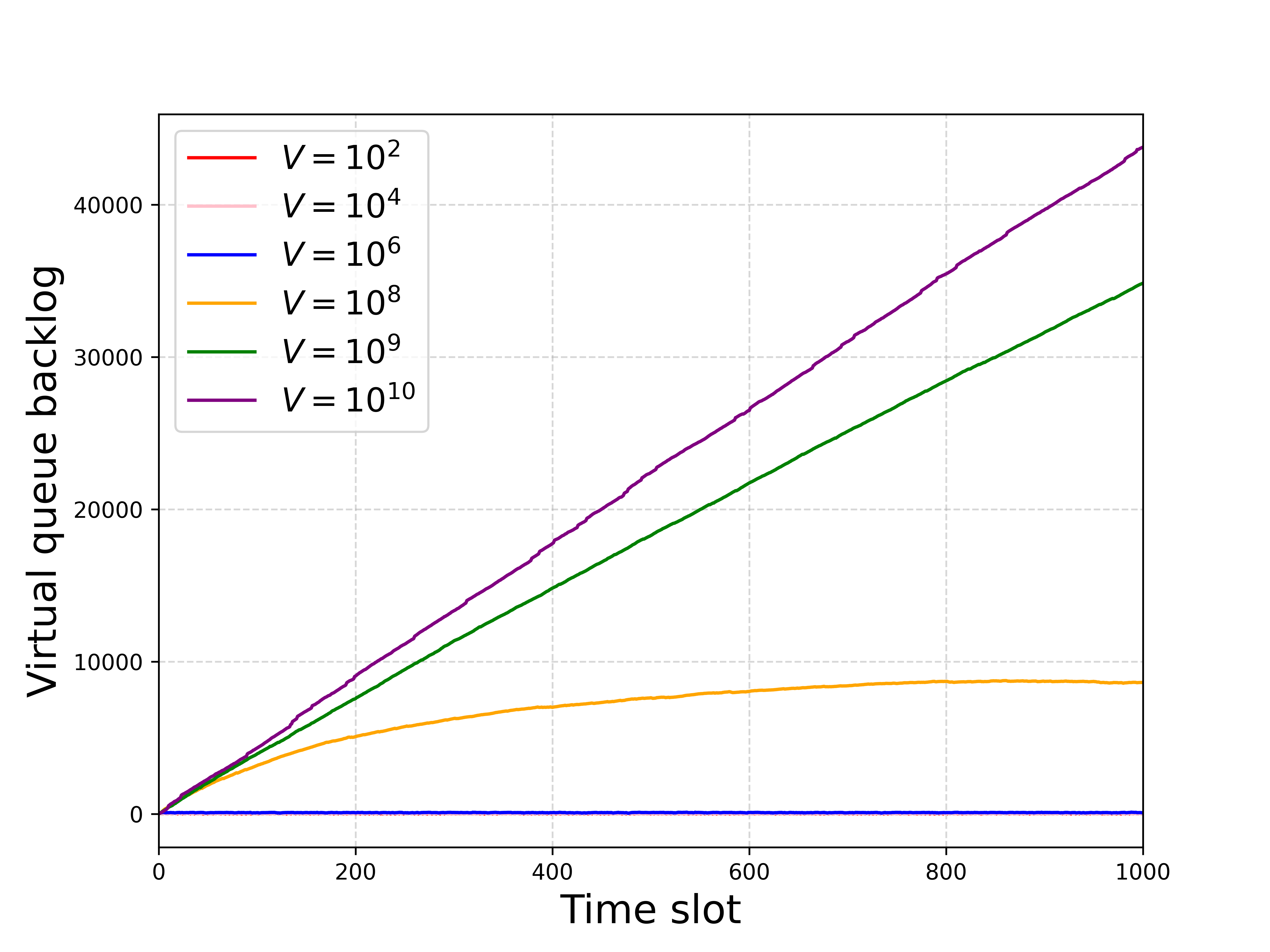}%
}
\subfloat[]{\includegraphics[width=0.33\textwidth]{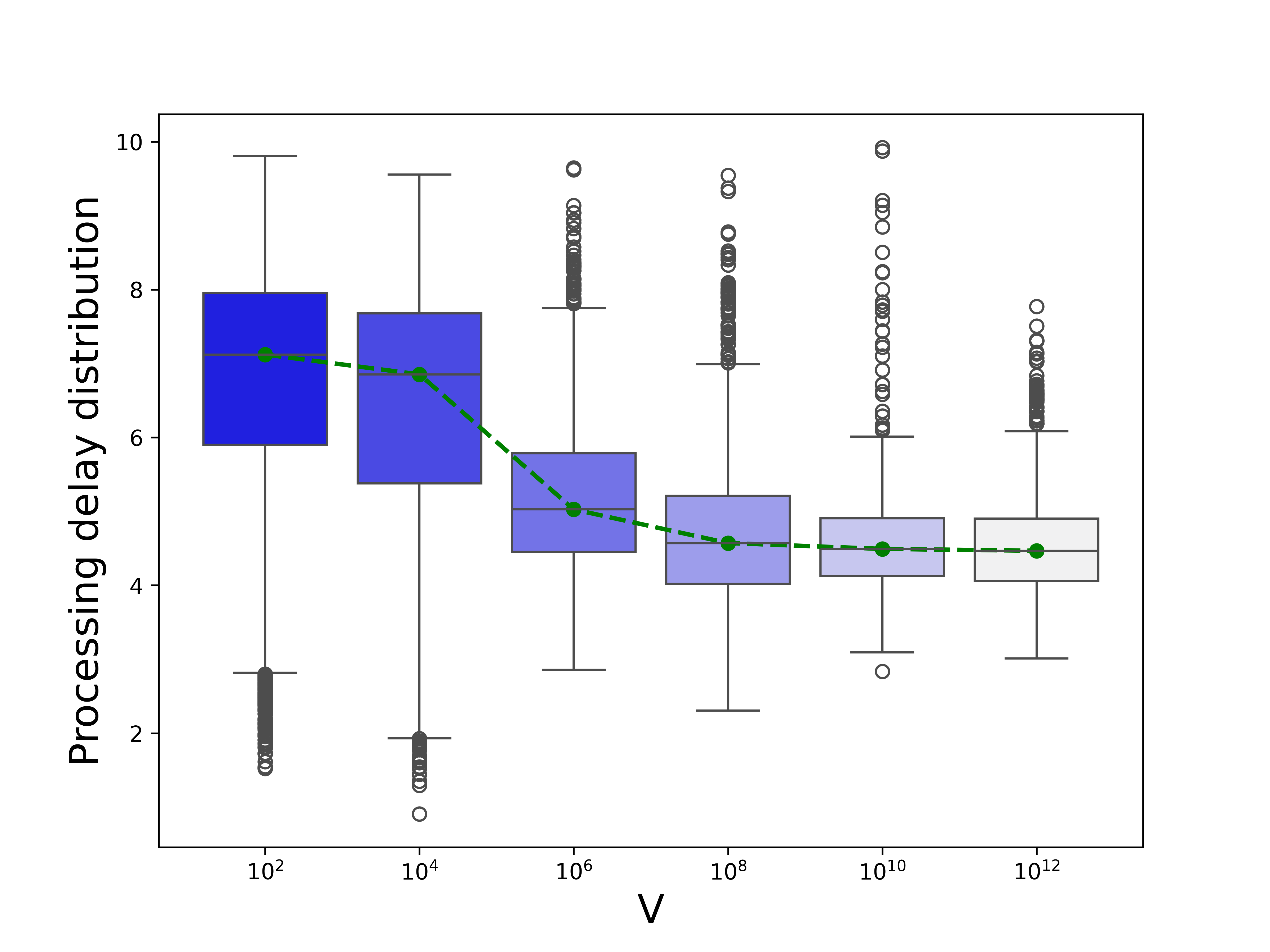}%
}
\vspace{-0.2cm}
\caption{Adaptive capability of the proposed scheme to the time-varying system state. (a) Variation of task processing cost over time for different values of $V$. (b) Variation of the virtual queue backlog over time for different values of $V$. (c) Distribution of task processing delays in each time slot for different values of $V$.}
\label{Adaptive capability}
\end{figure*}

\par Fig. \ref {Scalability Fig} compares the scalability of the two proposed schemes against the SH and DRL algorithms across multiple plants. The number of factories and the long-term average task processing cost budget are scaled by multiplying the default parameters by factors from 1 to 6. The resource allocation and request arrival settings for every additional five factories remain the same as those for the initial five factories. Fig. \ref {Scalability Fig} (a) presents the delay performance of different schemes. The average delay of the two proposed schemes and the SH algorithm gradually decreases as the number of factories increases. This is primarily because a higher number of factories enhances inter-factory collaboration, thereby reducing the average task processing delay. In contrast, the average delay of the DRL algorithm increases at larger $M$ as the number of factories grows. This is because the action space of the task scheduling policy expands quadratically with the number of factories, making it harder for the DRL algorithm to converge. At this stage, the DRL-generated task scheduling policy is largely random, leading to degraded delay performance. Fig. \ref {Scalability Fig} (b) presents the long-term average energy consumption of different schemes. The DRL scheme consistently underperforms compared to the Lyapunov-based scheme, primarily due to its low sample efficiency. It requires a sufficient number of samples before optimizing its strategy, which initially behaves as a random strategy. Additionally, the DRL scheme's performance heavily depends on the reward function design. Even a slight modification can cause convergence to a completely different strategy.

\par Fig. \ref{Scalability Fig}(c) presents the average running time of different algorithms versus the number of factories $M$. When the number of factories is small (e.g., $M = 5$), all algorithms exhibit similar average running times. However, as $M$ increases, the running times of the proposed HH algorithm and the SH algorithm grow significantly. Moreover, the proposed HH algorithm consistently exhibits a shorter running time than the SH algorithm. This is because the proposed HH algorithm reduces the search space for the optimal solution through node category assignment. While the running time of the proposed IL scheme increases with $M$, its growth is less significant than that of the proposed HH algorithm. This indicates that the search process for optimal category assignment is the primary factor contributing to the increased running time of the proposed HH algorithm. Additionally, although the DRL algorithm consistently exhibits the shortest running time, its delay and cost performance remain the worst, with a significant gap compared to the other three schemes. According to Fig. \ref{Performance comparison} and Fig. \ref{Scalability Fig}, the proposed HH scheme is preferable for smaller networks, and the proposed IL scheme is better suited for large-scale networks. This conclusion aligns with our theoretical analysis of the computational complexity of the two schemes.

\subsubsection{Adaptability to System State Changes}
\label{Adaptability to System State Changes}

\par Fig. \ref{Adaptive capability}(a) shows the task processing cost in each time slot. At smaller values of $V$, the task processing cost rapidly converges to the expected long-term average threshold. As $V$ increases, the convergence of the task processing cost curve gradually slows. For $V>10^8$, the task processing cost curve does not converge to the expected long-term average threshold. As $V$ increases further, the convergence value rises, convergence slows, and task processing cost fluctuations become more pronounced. This occurs because increasing $V$ reduces the emphasis on controlling energy consumption that exceeds the expected long-term average, weakening task processing cost control and causing the curve to fluctuate significantly or even fail to converge.

\par Fig. \ref{Adaptive capability}(b) shows the variation of the virtual queue backlog over time for different values of $V$. The virtual queue backlog remains near zero for $ V<10^8$. This occurs because the proposed scheme tightly controls task processing cost when $V$ is small, ensuring that processing energy consumption in each time slot does not exceed the long-term average task processing cost budget. At $V = 10^8$, the virtual queue backlog initially grows before stabilizing. This corresponds to Fig. \ref{Adaptive capability}(a), where the task processing cost gradually converges from a higher value to the long-term average task processing cost budget, after which the virtual queue backlog stabilizes. For $V > 10^8$, the virtual queue backlog does not converge and grows rapidly as $V$ increases, corresponding to the behavior in Fig. \ref{Adaptive capability}(a).

\par Fig. \ref{Adaptive capability}(c) shows the distribution of task processing delays across time slots for different values of $V$. As $V$ increases, the median task processing delay decreases. Additionally, for $ V > 10^4$, the distribution of task processing delays becomes more compact as $V$ increases. This occurs because as $V$ increases, the algorithm places more emphasis on minimizing delay, leading to tighter control over delay fluctuations across time slots.

\subsubsection{Impact of Resource Allocation and Request Distribution Match Degree on Algorithm Performance}
\label{SubSubSection: Impact of Resource Allocation and Request Distribution Match Degree on Algorithm Performance}

\par
\begin{figure}[t]
\centering
\subfloat[]{\includegraphics[width=0.25\textwidth]{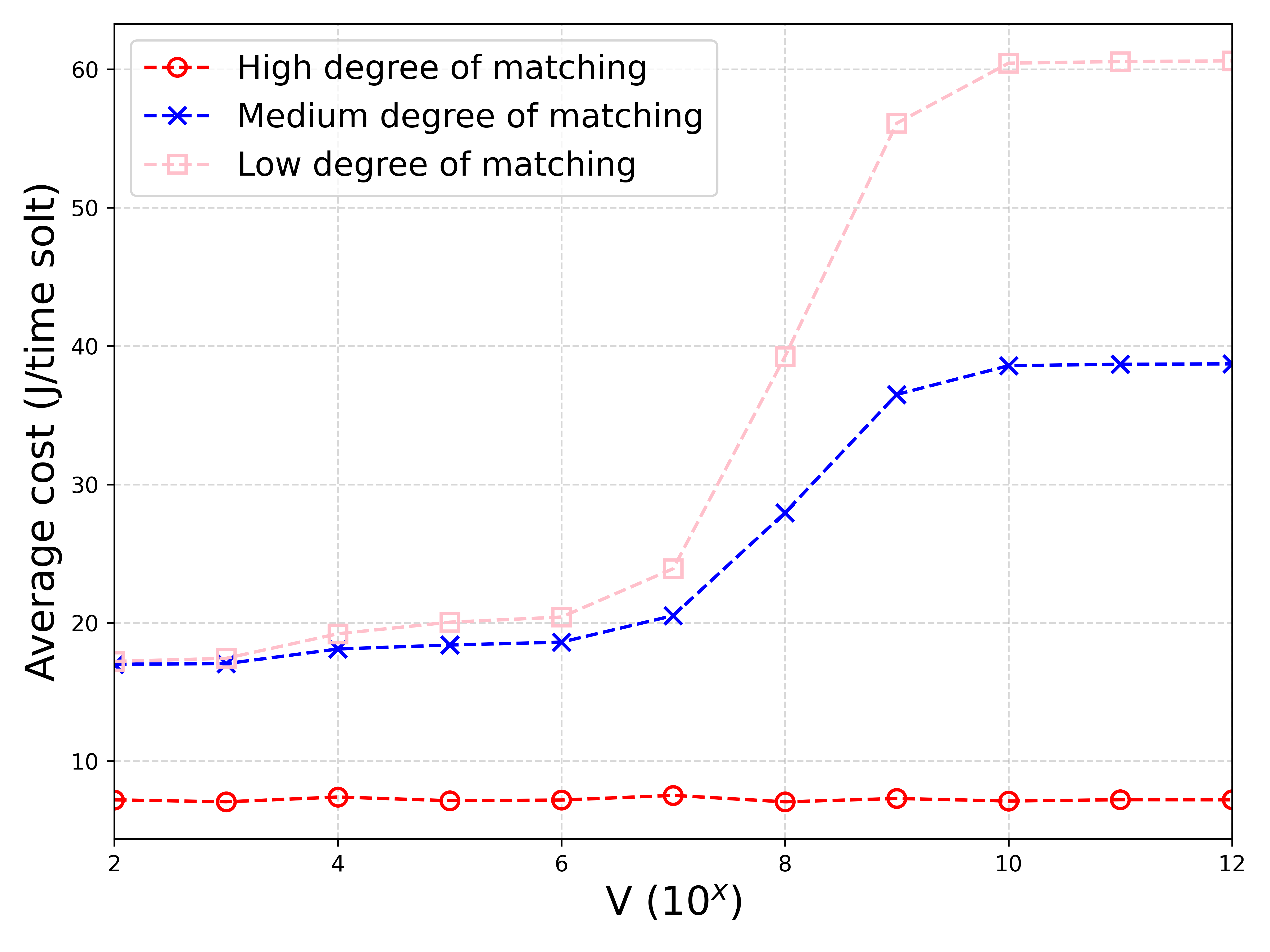}%
}
\subfloat[]{\includegraphics[width=0.25\textwidth]{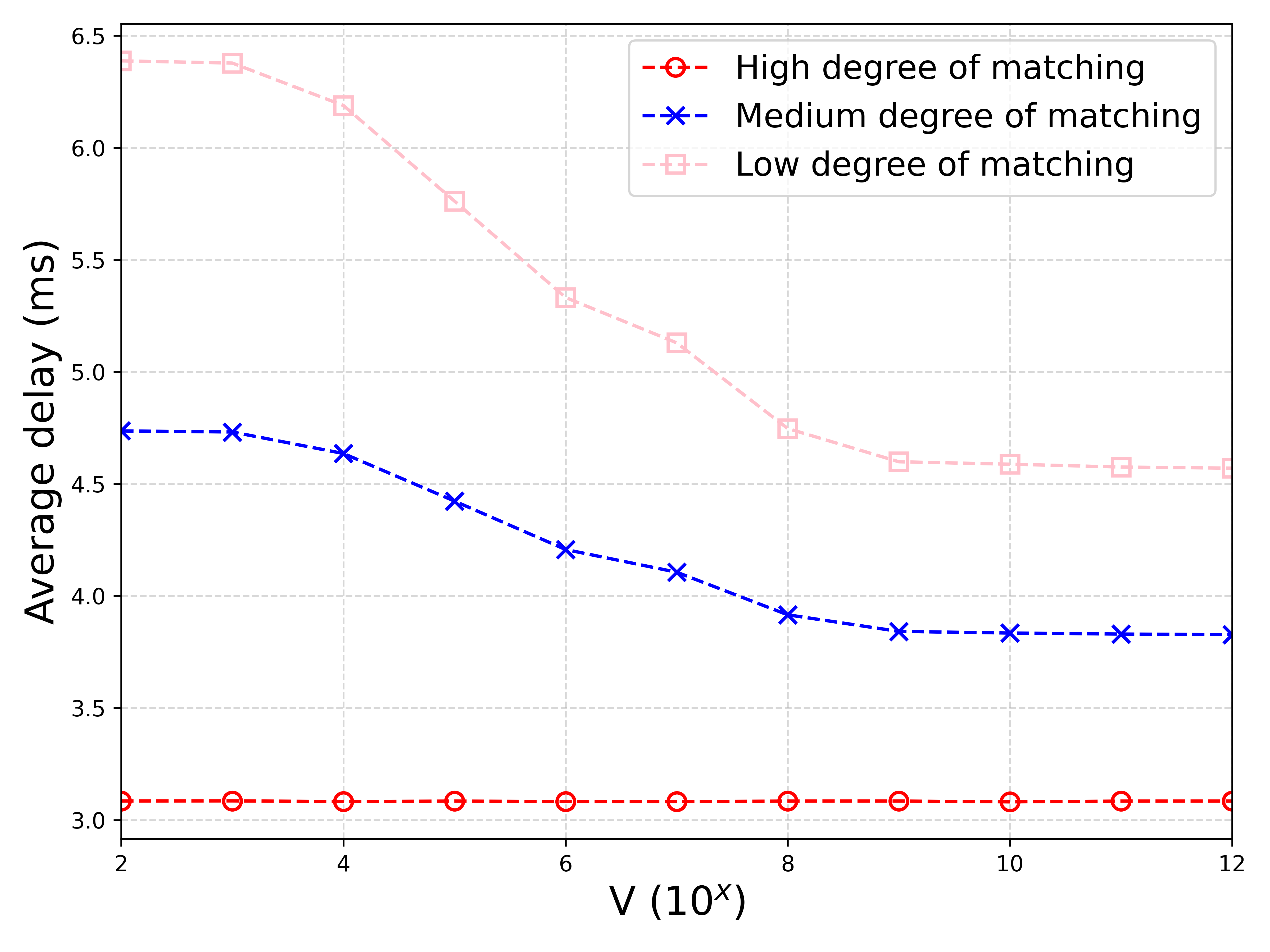}%
}
\vspace{-0.2cm}
\caption{Impact of the alignment between resource allocation and request distribution on algorithm performance. (a) The long-term average task processing energy consumption with different values of $V$. (b) The average response delay with different values of $V$.}
\label{matching}
\vspace{-0.2cm}
\end{figure}

 Fig. \ref{matching} illustrates the impact of the degree of match between resource allocation and request distribution on algorithm performance. For high matching, the ranges of values for $F_1^1(t)$ to $F_1^5(t)$ are [10, 20], [20, 30], [30, 40], [40, 50], and [50,60]; for medium matching, the ranges of values are [50, 60], [20, 30], [30, 40], [40, 50], and [10, 20]; for low matching, the ranges of values are [50, 60], [40, 50], [30, 40], [20, 30], and [10, 20], respectively. As shown in Fig. \ref{matching}(a) and (b), a low alignment between request distribution and resource allocation will lead to increased latency and higher task processing cost. The difference in processing cost becomes more significant as $V$ increases, while the impact of matching on average response delay is more pronounced at smaller values of $V$. Based on the above analysis, the degree of matching between resource allocation and task request distribution determines the upper bound of scheduling policy performance. As preliminary steps, service deployment and resource allocation play a dominant role in determining scheduling policy performance. Task scheduling aims to optimize long-term service quality for spatiotemporally varying requests, subject to a long-term average cost constraint, once service deployment and resource allocation are complete.

\subsubsection{Impact of the Long-Term Average Cost Budget on Algorithm Performance}
\label{SubSubSection: Impact of the Long-Term Average Cost Budget on Algorithm Performance}

\begin{figure}[t]
\vspace{-0.2cm}
\centering
\subfloat[]{\includegraphics[width=0.25\textwidth]{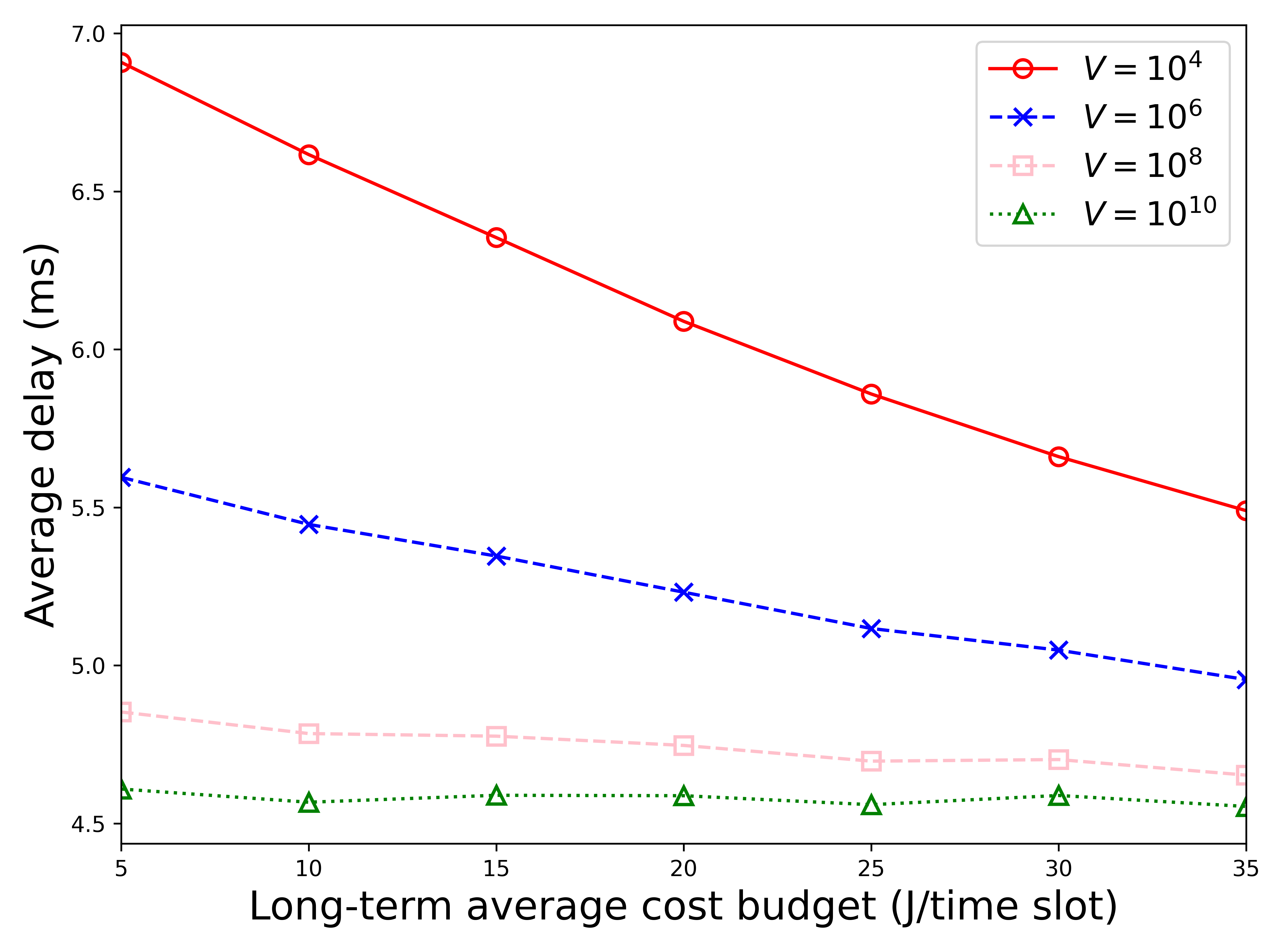}%
}
\subfloat[]{\includegraphics[width=0.25\textwidth]{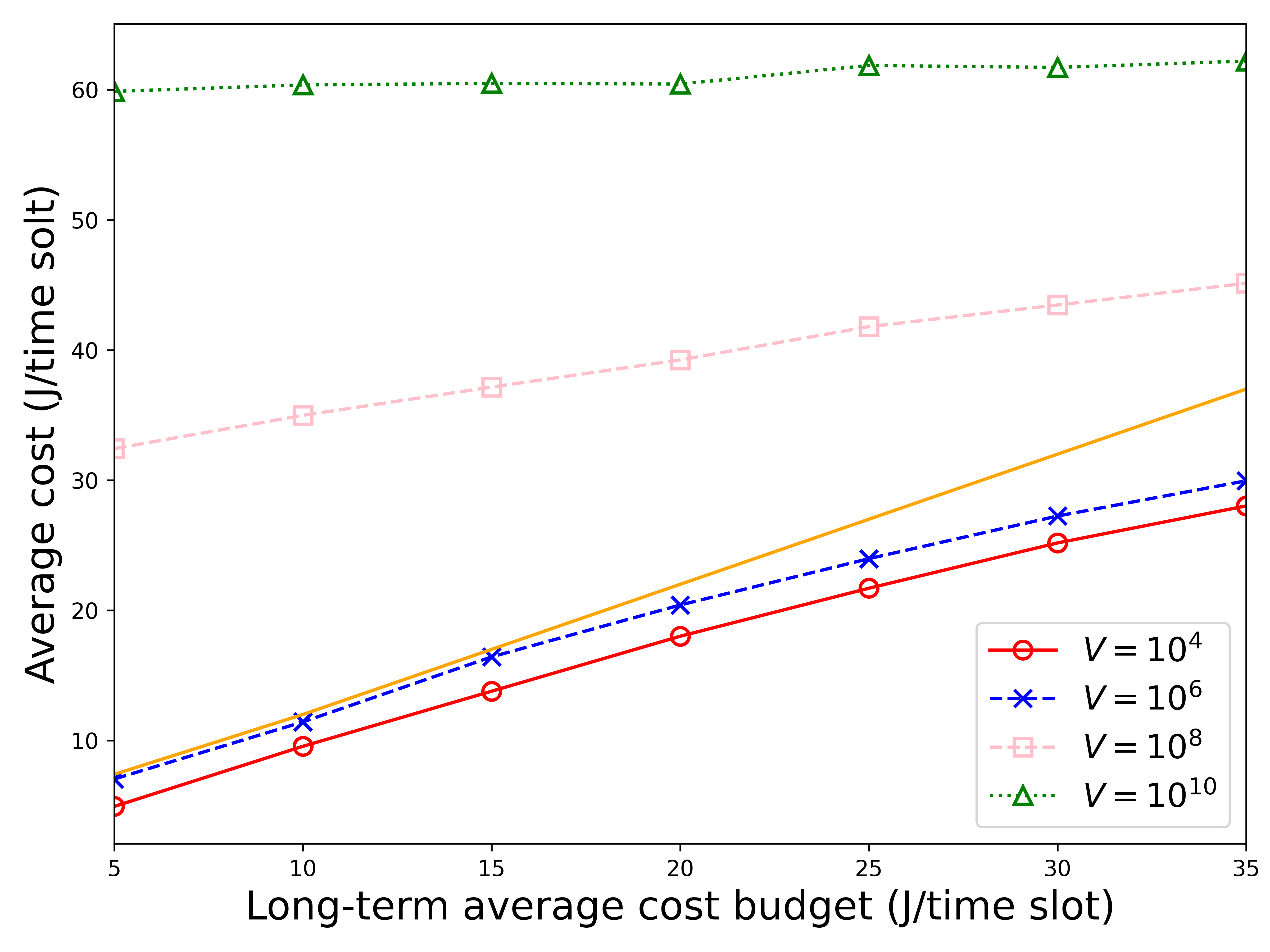}%
}
\vspace{-0.2cm}
\caption{Impact of long-term average cost constraint on algorithm performance. (a) Variation in average response latency with the long-term average cost constraint. (b) Variation in average processing energy consumption with the long-term average cost constraint.}
\label{constraint}
\end{figure}

\par Fig. \ref{constraint} illustrates the variation in delay and cost performance of the proposed scheme under different values of $V$ and the long-term average cost constraint. At very large values of $V$, both delay and cost remain nearly constant. For other values of $V$, as the long-term average cost constraint is relaxed, the proposed scheme achieves a lower average task processing delay while average task processing energy consumption increases gradually. This occurs because relaxing the long-term average energy consumption constraint increases the cost budget for task processing, allowing the more optimal solution to be found over a larger search space. The orange solid line in Fig. \ref{constraint}(b) represents the long-term average energy consumption expectation. It can be seen that the proposed algorithm minimizes delay performance only when $V < 10^6$ while strictly ensuring that the long-term average energy consumption constraint is not exceeded.

\subsection{Practicability Evaluation}
\label{SubSection:Practicability Evaluation}

\begin{figure}[h]
    \centerline{\includegraphics[width=0.9\linewidth]{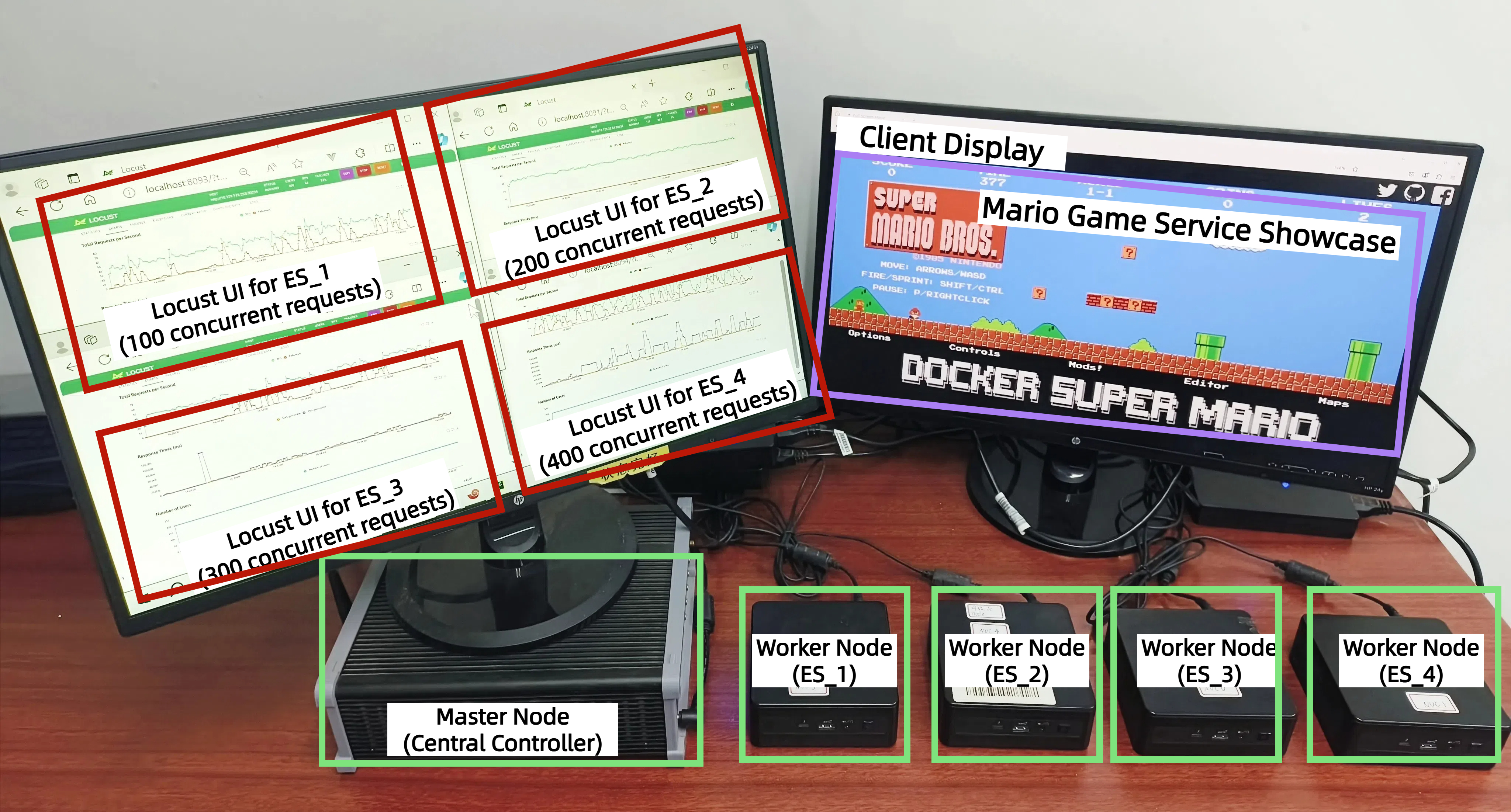}}
    \caption{Testbed for our proposed task scheduling scheme.}
    \label{v-c1}
\end{figure}
\begin{figure}[h]
    \centerline{\includegraphics[width=0.9\linewidth]{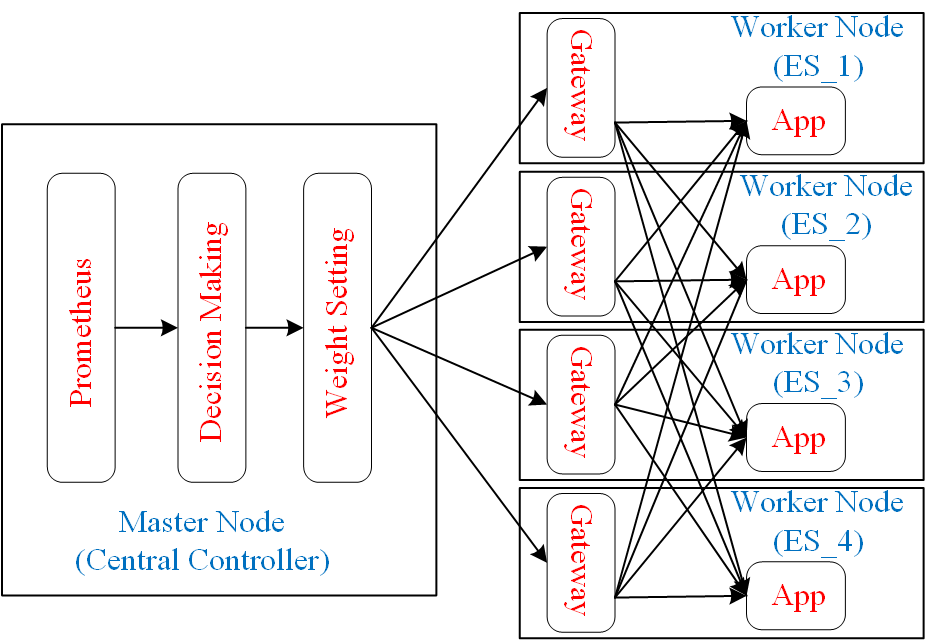}}
    \caption{System Architecture.}
    \label{v-c2}
\end{figure}

\par In this section, we assess the practicality of the proposed scheme. As shown in Fig. \ref{v-c1}, we build a Kubernetes-based testbed, consisting of four Intel minicomputer hosts as worker nodes (representing the four ESs integrated with the APs in Fig. \ref{Figure:System Model}) and one GeeFlex host as the master node (corresponding to the central controller). Each worker node deploys a gateway (corresponding to the AP in Fig. \ref{Figure:System Model}) and an application pod (Mario's cloud gaming service). We allocate 2 CPU cores and 2 GB of RAM to each service pod by configuring the Kubernetes resource file. Using Istio, an open-source service mesh, we configure the gateway of each worker node to distribute incoming user requests to the application pods based on predefined weights \cite{35}. Our algorithm is used to adjust the weight (scheduling probability) of each gateway to each service pod, achieving task scheduling across nodes. As shown in Fig. \ref{v-c2}, the algorithm module is deployed on the master node. The Prometheus module monitors the system state, while the Decision Making module periodically reads the system state from the Prometheus module and runs the algorithm (Due to practical difficulties in measuring energy consumption from task scheduling, we simplify the evaluation by considering only response latency.). The decision result, i.e., the weight of each gateway to each service pod, is generated and transmitted to the Weight Setting module. The Weight Setting module modifies the weight of each gateway to each service pod based on the decision from the Decision Making module.

\par To validate the performance of our scheme, we use Locust, a Python-based performance testing framework, to send varying numbers of concurrent requests to each worker node's gateway, simulating spatial non-uniformity in user request distribution. In our experiments, the numbers of concurrent requests sent to the four gateways are set to 100, 200, 300, and 400, with the average response latency of all requests as the observation metric.

\par We compare our proposed scheme with task scheduling strategies frequently used in engineering practice: round-robin scheduling, load-based scheduling, least-connection scheduling, and direct processing. The comparison result in Table \ref{tab2} shows that our proposed scheme achieves lower task response latency. This is primarily because our scheme unifies the consideration of both computing power and network states. The average task response latency under our scheme is reduced by approximately $20\%$ compared to the load-based task scheduling mechanism.

\begin{table}[h]
\centering
\caption{Average Response Latency Comparision of the Five Algorithms
}
\label{tab2}
\resizebox{\columnwidth}{!}{%
\begin{tabular}{l|l|l}
\hline
\textbf{Scheme}                                                     & \textbf{Feature}                                                                                                                    & \textbf{Latency} \\ \hline
Proposed Scheme                                                     & \begin{tabular}[c]{@{}l@{}}Introduce a graph model to guide\\ optimal task scheduling decisions\end{tabular} & 33 ms            \\ \hline
Round-robin Scheduling                                                 & \begin{tabular}[c]{@{}l@{}}Distribute requests to available\\ server nodes sequentially and \\ cyclically.\end{tabular}             & 49 ms            \\ \hline
Load-based Scheduling                                                  & \begin{tabular}[c]{@{}l@{}}Distribute requests according to\\ each server's current load.\end{tabular}                              & 41 ms            \\ \hline
\begin{tabular}[c]{@{}l@{}}Least-connection \\ Scheduling\end{tabular} & \begin{tabular}[c]{@{}l@{}}Prioritize scheduling requests to\\ servers with the fewest active \\ connections.\end{tabular}          & 43 ms            \\ \hline
Direct Processing                                                      & \begin{tabular}[c]{@{}l@{}}Each edge node directly handles\\ incoming requests without \\scheduling.\end{tabular}                     & 62 ms            \\ \hline
\end{tabular}%
}
\end{table}
\section{Conclusion}
\label{Section:Conclusion}

\par In this paper, we address the problem of optimizing the performance of collaborative edge network services under a long-term average task processing cost budget. We implement a spatiotemporal non-uniformity-aware online task scheduling framework to balance average response delay and task scheduling cost. To handle unpredictable future system states, we incorporate the long-term budget into real-time optimization problems using Lyapunov optimization techniques. To address the spatial non-uniformity of user request distribution, we introduce a graph model and derive the conditions for the optimal solution. We then employ an enhanced discrete particle swarm algorithm and a harmonic search algorithm to get the near-optimal solution based on the derived conditions. To accelerate the algorithm’s operation further, an imitation-based approach is introduced. We also provide a theoretical analysis of the algorithm's performance. Finally, we conduct extensive simulations to demonstrate the effectiveness of the proposed algorithms and build a prototype testbed to validate the practicality of the proposed scheme.

\bibliographystyle{IEEEtran}
\bibliography{cite}

\begin{IEEEbiography}[{\includegraphics[width=1in,height=1.25in,clip,keepaspectratio]{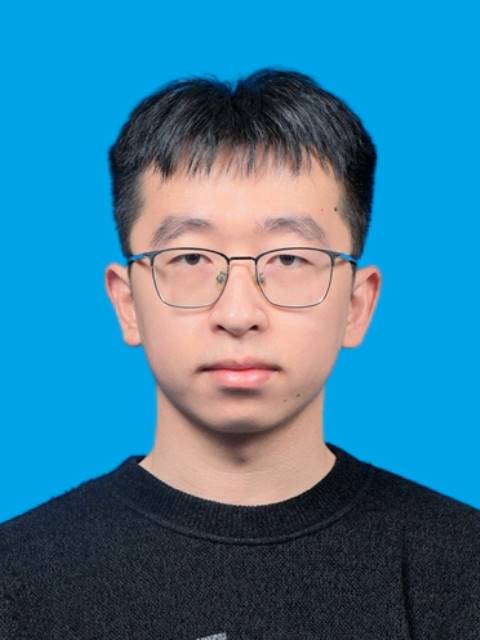}}]{Yang Li} received the B.S. degree in communication engineering from Beijing University of Posts and Telecommunications, Beijing, China, in 2022, where he is currently pursuing the Ph.D. degree with the Key Laboratory of Universal Wireless Communication, School of Information and Communication Engineering.
\par His research interests include mobile edge networks, Internet of Things, and resource allocation.
\end{IEEEbiography}

\begin{IEEEbiography}[{\includegraphics[width=1in,height=1.25in,clip,keepaspectratio]{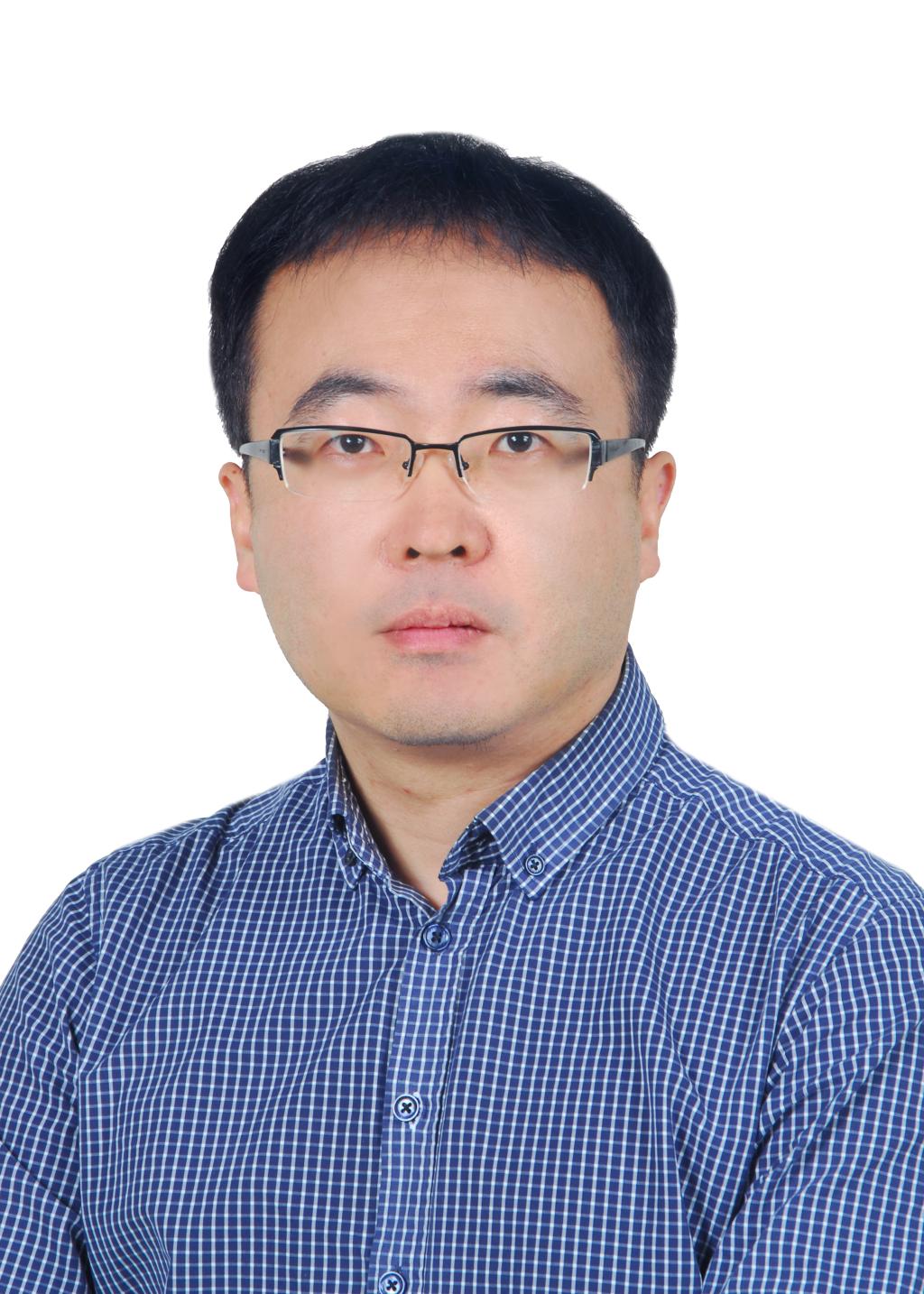}}]{Xing Zhang} (Senior Member, IEEE) received the Ph.D. degree from Beijing University of Posts and Telecommunications (BUPT), Beijing, China, in 2007.
\par He is currently a Full Professor with the School of Information and Communications Engineering, BUPT. He has authored or coauthored five technical books and over 300 papers in top journals and international conferences and holds over 80 patents. His research interests are mainly in 5G/6G networks, satellite communications, edge intelligence, and Internet of Things.
\par Dr. Zhang has received six Best Paper Awards in international conferences. He has served as a General Co Chair for the Third IEEE International Conference on Smart Data (SmartData-2017) and as a TPC co chair/TPC member for a number of major international conferences. 
\end{IEEEbiography}

\begin{IEEEbiography}[{\includegraphics[width=1in,height=1.5in,clip,keepaspectratio]{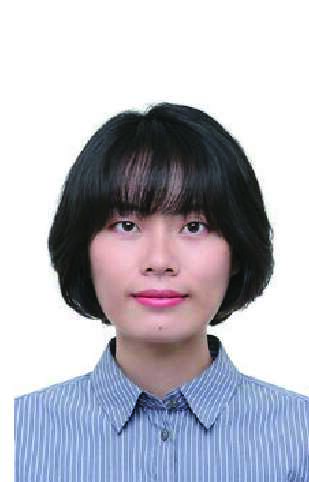}}]{Yukun Sun} received the B.S. degree in communication engineering from Beijing University of Posts and Telecommunications (BUPT), Beijing, China, in 2020. She is currently pursuing the Ph.D.degree with the Key Laboratory of Universal Wireless Communication, School of Information and Communication Engineering, BUPT. 
\par Her research interests include computing power network, computing offloading and resource allocation.
\end{IEEEbiography}

\begin{IEEEbiography}[{\includegraphics[width=1in,height=1.25in,clip,keepaspectratio]{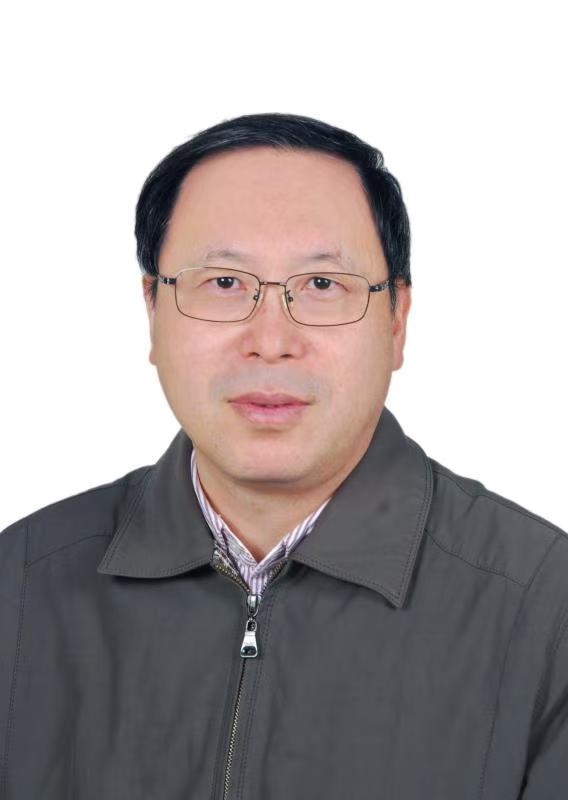}}]{Wenbo Wang} (Senior Member, IEEE) received the B.S., M.S., and Ph.D. degrees from Beijing University of Posts and Telecommunications (BUPT), Beijing, China, in 1986, 1989, and 1992, respectively.
\par He is a former Vice President and a Professor at BUPT, a Fellow of the China Institute of Communications. He is currently the President of the Xiong’an Institute of Aerospace Information, Supervisor of the Chinese Association for Artificial Intelligence, Chairman of the Academic Work Committee of the China Institute of Communications, and Chairman of the Digital Twin and System Simulation Professional Committee of the China Institute of Communications. His current research interests include radio transmission technology, wireless network theory, cognitive communications, and software radio technology.
\end{IEEEbiography}

\begin{IEEEbiography}[{\includegraphics[width=1in,height=1.25in,clip,keepaspectratio]{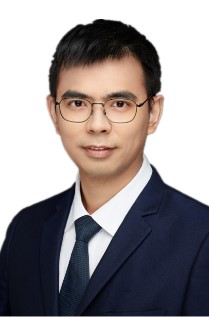}}]{Bo Lei} received the master’s degree in telecommunication engineering from Beijing University of Posts and Telecommunications, Beijing, China, in 2006.
\par He is now the deputy director of the Network Technology Research Institute of a director of future network research center of China telecom research institute. He currently leads the Future Network Research Center focusing on future network. He is the first author of two technical books and has published more than 30 papers in top journals and international conferences, and filed more than 30 patents. His currently research interests include future network architecture, new network technology, computing power network, and 5G application verification.
\end{IEEEbiography}

\vfill

\end{document}